\documentclass{aa}

\usepackage{silence}
\usepackage{graphicx}
\usepackage{txfonts}
\usepackage{array}
\usepackage{siunitx}
\usepackage{pdflscape} 
\usepackage{multicol}
\usepackage{booktabs}
\usepackage{xcolor}

\vbadness=\maxdimen

\usepackage{color}
\usepackage{natbib,twoopt}
\usepackage[breaklinks]{hyperref}      
\bibpunct{(}{)}{;}{a}{}{,}             
\hypersetup{
  colorlinks,
  citecolor=blue,
  linkcolor=blue,
  urlcolor=blue,
}
\makeatletter
  \newcommandtwoopt{\citeads}[3][][]{\href{http://adsabs.harvard.edu/abs/#3}%
    {\def\hyper@linkstart##1##2{}%
     \let\hyper@linkend\@empty\citealp[#1][#2]{#3}}}
  \newcommandtwoopt{\citepads}[3][][]{\href{http://adsabs.harvard.edu/abs/#3}%
    {\def\hyper@linkstart##1##2{}%
     \let\hyper@linkend\@empty\citep[#1][#2]{#3}}}
  \newcommandtwoopt{\citetads}[3][][]{\href{http://adsabs.harvard.edu/abs/#3}%
    {\def\hyper@linkstart##1##2{}%
     \let\hyper@linkend\@empty\citet[#1][#2]{#3}}}
  \newcommandtwoopt{\citeyearads}[3][][]%
    {\href{http://adsabs.harvard.edu/abs/#3}
    {\def\hyper@linkstart##1##2{}%
     \let\hyper@linkend\@empty\citeyear[#1][#2]{#3}}}
\makeatother

\newcolumntype{M}[1]{>{\raggedright \arraybackslash}m{#1}} 
\newcolumntype{C}[1]{>{\centering \arraybackslash}m{#1}} 
\newcolumntype{R}[1]{>{\raggedright\arraybackslash}p{#1}} 

\newcolumntype{f}[1]{>{\footnotesize \raggedright \arraybackslash}m{#1}} 

\newcolumntype{B}{>{\bfseries \boldmath}l} 

\graphicspath{{./images/}}

\DeclareSIUnit\angstrom{\text {Å}}

\begin{document}

   \title{Unveiling the dynamics of the ultra-fast outflow in IRAS 13224-3809 with X-ray spectroscopy}

   \subtitle{}

   \author{Pierpaolo Condò \thanks{pierpaolo.condo@inaf.it}
          \inst{1,2,3}
          \and
          Francesco Tombesi
          \inst{1,2,3}
          \and
          Marco Laurenti\inst{1,2}
          \and
          Alfredo Luminari\inst{3,4}
          \and
          Riccardo Middei\inst{3,5}
          \and
          Enrico Piconcelli\inst{3}
          \and
          Massimo Gaspari\inst{6}
          \and
          Giorgio Lanzuisi\inst{7}
          \and
          Roberto Serafinelli\inst{8,3}
          \and
          Alessia Tortosa\inst{3}
          \and
          Luca Zappacosta\inst{3}
          \and
          Fabrizio Nicastro\inst{3}
          }

   \institute{
    Dipartimento di Fisica, Università di Roma Tor Vergata, Via della Ricerca Scientifica, 1, Roma 00133, Italy
    \and
    INFN - Rome Tor Vergata, Via della Ricerca Scientifica 1,
    00133 Rome, Italy
    \and
    INAF - Osservatorio Astronomico di Roma, Via Frascati 33, I-00040 Monte Porzio Catone, Italy
    \and
    INAF - Istituto di Astrofisica e Planetologia Spaziali, Via del Fosso del Cavaliere, 100, Roma 00133, Italy
    \and
    Space Science Data Center, Agenzia Spaziale Italiana, Via del Politecnico snc, 00133 Roma, Italy
    \and
    Department of Physics, Informatics and Mathematics, University of Modena and Reggio Emilia, 41125 Modena, Italy
    \and
    INAF $-$ Osservatorio di Astrofisica e Scienza dello Spazio di Bologna, Via Gobetti 101, I-40129 Bologna, Italy
    \and
    Instituto de Estudios Astrofísicos, Facultad de Ingeniería y Ciencias, Universidad Diego Portales, Avenida Ejército Libertador 441, Santiago, Chile
    }

   \date{Received ...; accepted ...}

\abstract{
IRAS 13224-3809 is one of the most intensively studied narrow-line Seyfert 1 galaxies, with a rich literature reporting diverse and sometimes contrasting interpretations of its complex X-ray spectra and variability. Notably, a fast and variable ultra-fast outflow (UFO) was discovered in this source, sparking debate over its nature and driving mechanisms. Motivated by these open questions, we present a systematic, time- and flux-resolved reanalysis of the full 2016 XMM-Newton (1.5\,Ms) and NuSTAR (500\,ks) datasets, employing careful background treatment and equal-count spectral selections.
We uniformly apply three spectral models, including photo-ionized absorption, broad emission, and relativistic reflection, to all intervals. We unambiguously confirm the presence of a strong, variable outflow with velocities exceeding $0.2c$, and find that models including absorption consistently reveal robust physical trends: a velocity-luminosity correlation of the UFO, persistently large line widths, and no compelling equivalent-width-flux anti-correlation. When emission or reflection components are included, the significance of the absorption features decreases, but significant UFO detections remain in most intervals.
We also report clear evidence for rapid acceleration of the wind in response to X-ray flares, with the outflow carrying momentum and kinetic power sufficient to drive an efficient AGN feedback. The observed rapid response favors magnetic driving, analogous to coronal mass ejections, over radiative acceleration. Our results reconcile contrasting previous claims and underline the need for high-resolution spectroscopy to resolve the wind substructure. The observed UFO variability and structure are consistent with a multiphase, clumpy wind produced by thermal and hydrodynamic instabilities, with magnetic reconnection providing the rapid acceleration mechanism.
}

\keywords{Galaxies: active, Galaxies: Seyfert: individual: IRAS 13224-3809}

\maketitle
\titlerunning
\authorrunning

\section{Introduction}
Active Galactic Nuclei (AGNs) are primarily powered by accretion onto supermassive black holes (SMBHs) and host some of the most energetic phenomena in the universe. Among the key observational signatures of AGN activity are ultra-fast outflows (UFOs), detected as blue-shifted Fe XXV/XXVI absorption lines above 7 keV in X-ray spectra, indicating velocities that can reach a substantial fraction of the speed of light \citep{pounds_evidence_2003, tombesi_evidence_2011}. These outflows are thought to originate from the inner regions of the accretion disk, within a few hundred gravitational radii of the black hole \citep{kraemer_physical_2018}, and are significant both for understanding accretion physics and their potential role in AGN feedback \citep{king_powerful_2015, gaspari_windrev_2020}.\\
Systematic surveys of nearby Seyfert galaxies have revealed that approximately 40\% of these AGNs harbor highly ionized UFOs, characterized by velocities exceeding 10 000 km s$^{-1}$ and averaging between 0.1$c$ and 0.3$c$ \citep{tombesi_evidence_2010, gofford_suzaku_2013, matzeu_subways_2023}. Various theoretical models have been proposed to explain the origin and structure of UFOs. These include radiation-driven winds from the accretion disk, magnetically-driven outflows, and multiphase cycles such as Chaotic Cold Accretion (CCA), where cold clouds condense from the hot halo and interact with the disk \citep{gaspari_cca_2013, gaspari_raining_2017}. Such mechanisms could naturally produce clumpy and transient outflows rather than smooth and continuous flows \citep{gaspari_unifying_2017, wittor_dissecting_2020}.\\
Recent groundbreaking results from the XRISM telescope have revolutionized our understanding of ultra-fast outflows. High-resolution spectroscopy has revealed that what appeared as single broad absorption features in CCD spectra are, in fact, complex superpositions of multiple narrow absorption lines \citep{xrism_pdsNature_2025, noda_iras_2025, mehdipour_ngc3783_2025, gu_ngc3783_2025, mizumoto_pg1211_2025,reeves_ngc4051_2026}. These "forests" of velocity-resolved components indicate that UFOs are highly structured and clumpy rather than homogeneous flows. However, despite this progress, several key questions remain: the launching mechanism, acceleration, and composition of winds are still poorly constrained. Detailed time-resolved studies of individual sources are therefore crucial to understand the dynamics and evolution of these powerful phenomena.\\

\section{IRAS 13224-3809}
A major advancement in UFO studies was made by \citet{parker_response_2017} with the identification of a variable, mildly-relativistic wind in the bright Narrow-Line Seyfert 1 (NLS1) galaxy IRAS 13224-3809 ($z = 0.0658$), based on extensive XMM-Newton observations in 2016 (1.5 Ms). Multiple absorption features were detected, with a significant line at $E=8.6$ keV indicating outflow velocities of $\upsilon = (0.236\pm 0.006)\, c$. Notably, Fe K absorption features weaken as X-ray luminosity increases, interpreted as over-ionization of the gas. A direct correlation was observed between UFO velocity and source luminosity, along with flux-dependent trends in equivalent width, ionization state, and column density \citep{parker_response_2017, pinto_ultrafast_2018}.\\
However, \citet{chartas_variable_2018} raised important caveats. After reanalyzing with optimized extraction regions to reduce background contamination above 7 keV, they found the flux-dependent trends were artifacts: only the velocity-luminosity correlation survived. This might suggest radiative acceleration \citep{sadowski_kinetic_2017}, though contrasting results in 1H 0707-495 (velocity-luminosity anti-correlation) complicate interpretation \citep{xu_wind-luminosity_2021}. These CCD-based conclusions remain highly degenerate, and the source's extreme variability (factor of 10 in $F_{\rm{0.3-10\,keV}}$ within $<$10 ks) demands careful spectral selection.\\
Despite extensive work by \citet{parker_response_2017}, \citet{pinto_ultrafast_2018}, \citet{chartas_variable_2018}, \citet{jiang_15_2018}, \citet{jiang_xmm-newton_2022}, and \citet{midooka_radiatively_2023}, among others, there is still no unified picture of IRAS 13224-3809's complex X-ray spectrum. Different models have been proposed: the hard spectrum (2-10 keV) has been modeled as either a direct power law or a relativistically reflected continuum; the soft excess (0.3-2 keV) has been attributed to multiple reflection components, or to variable partial covering by clumpy absorbers. The persistent model degeneracy underscores the need for a comprehensive, systematic approach.\\
This work presents a systematic reanalysis of the full 2016 XMM-Newton (1.5 Ms) and NuSTAR (500 ks) campaign, with careful background treatment and equal-count spectral selections in both flux and time domains. We uniformly apply three alternative spectral models across all intervals, allowing us to track the evolution of UFO parameters robustly. A key novelty of our approach is the ability to measure the wind acceleration in response to X-ray flares, a direct probe of the driving mechanism that has not been systematically explored in previous studies. We focus on the hard X-ray band ($E>3$ keV) where UFO features are most prominent and previous claims are based \citep{parker_response_2017, chartas_variable_2018}.

In Section \ref{sec:red} we report on the data reduction. Sections \ref{sec:average} and \ref{sec:LMF_specs} are dedicated to the spectral modeling while in Sections \ref{sec:discussion} and \ref{sec:conclusions} we discuss the results and draw the conclusions.

Throughout this paper, unless otherwise specified, errors are at the 68\% ($\sim1\sigma$ for Gaussian distributions) confidence level, and the upper and lower limits are at the 90\% confidence level.
The reported errors on the best-fit parameters are calculated using the likelihood profiling method through the \texttt{error} command in \texttt{XSPEC}.


\begin{figure*}
    \centering
    \includegraphics{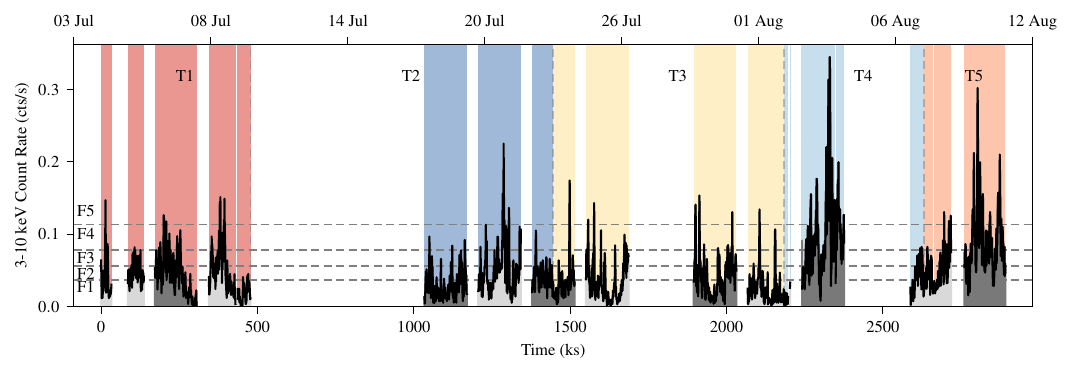}
    \caption{XMM-Newton EPIC-pn light curves of IRAS 13224-3809 in the 3--10~keV band. The plot shows the division of the data into five time intervals (T1--T5, labeled at the top) and five flux intervals (F1--F5, labeled on the left), each selected to contain equal total counts for optimal signal-to-noise in subsequent spectral analysis. Vertical dashed lines mark the boundaries of the time intervals, while horizontal dashed lines indicate the flux thresholds. The flux thresholds used are $F = 0.037, 0.056, 0.078, 0.114$ cts s$^{-1}$.
    The bin size of the light curves is 1000 ks.}
    \label{fig:lc}
\end{figure*}
\section{Data Reduction and Products Extraction}
\label{sec:red}

We use the full 1.5 Ms ($\sim$17 days) observing campaign with the X-ray Multi-Mirror Mission (XMM-Newton; \citet{StruderXMM}) and 500 ks (almost 6 days) with the Nuclear Spectroscopic Telescope Array (NuSTAR) high-energy X-ray mission \citep{Harrison2013} spanning between July and August 2016.\\
Observation dates, exposure times, and background-subtracted source count rates are listed in Table~\ref{tab:obs}. All energies are quoted in the source rest frame at $z=0.0658$ \citep{6dFSurvey}. Spectral fitting was performed with the X-ray spectral fitting package \texttt{XSPEC} v12.15.1 \citep{XSPEC}. All models include Galactic absorption via \texttt{TBabs} ($N_H = 4.76 \times 10^{20}$ cm$^{-2}$; \citealt{HI4PI_2016}). To account for cross-calibration between the XMM-Newton EPIC-pn and the NuSTAR focal plane modules (FPMs), we included a multiplicative constant free to vary between detectors; its value remains within 25\% for all fits including NuSTAR data, consistent with calibration uncertainties given the non-simultaneity of the observations and the source variability.\\

\subsection{XMM-Newton Data}
We limited the analysis to the European Photon Imaging Camera pn (EPIC-pn) data to maximize the signal-to-noise ratio in the hard X-ray band, as this instrument has a greater effective area than the two combined MOS detectors between 0.3 and 10 keV. For the reduction, we filtered the EPIC-pn data by selecting events corresponding to instrument \texttt{PATTERNS} in the 0-4 range (single and double pixel events). Several moderate-amplitude background flares were present during several of the XMM-Newton observations and were filtered out by excluding time intervals with rate higher than 0.4 counts/s in the energy range 10-12 keV.
To test sensitivity to background non-uniformity, we varied the position of the circular background extraction region (fixed radius of 50 arcseconds) across distances of 100--300 arcseconds from the source on the same CCD chip, finding no differences in spectral shapes or features.\\
Before proceeding to the flux- and time-resolved extraction of EPIC-pn spectra, we systematically tested different source extraction regions to maximize the signal-to-noise ratio in the detector-frame 7--10~keV band, where UFO features are expected. \citet{chartas_variable_2018} demonstrated that background contamination above $\sim$7~keV is significant for the larger extraction regions used in \citet{parker_response_2017}, and that selecting smaller apertures (250 physical units, or 12.5~arcsec) improves the signal-to-noise ratio by a factor of $\sim$8 by reducing background contribution while retaining $\sim$60\% of the XMM-Newton point spread function.
To reproduce these findings, we tested circular source extraction regions with radii of 30, 25, 20, 15, 12.5, and 10~arcseconds on two representative observations (Obs.~ID 0780561601 and 0792180201), comparing the 7--10~keV signal-to-noise ratio for each aperture. For the background, we maintained the fixed circular extraction region radius of 50 arcseconds on the same chip as the source.
Consistent with \citet{chartas_variable_2018}, we confirm that a 12.5-arcsecond radius optimizes the hard X-ray signal-to-noise and adopted this extraction region for all subsequent analysis.
The extraction of the products was performed using version 21.0 of the XMM-Newton Science Analysis System (SAS) and following the XMM-Newton
\href{https://www.cosmos.esa.int/web/xmm-newton/sas-threads}{SAS Threads}. Spectra and light curves were extracted with the \texttt{evselect} task. For each extraction, we generated the ancillary files and response matrices using, respectively, the \texttt{rmfgen} and \texttt{arfgen} tasks.\\ 
To obtain an average spectrum with a high signal-to-noise ratio, we stacked all EPIC-pn exposures using the \texttt{epicspeccombine} task. In this case, signal-to-noise is not a problem, and we binned the spectrum to not oversample the instrumental energy resolution by more than a factor of three. Simply with this criterion, the average spectrum already has more than 25 counts per bin in the 3--10~keV band.\\
To complement the X-ray dataset, we extracted UV data from the Optical Monitor (OM) onboard XMM-Newton, which observed IRAS 13224-3809 in imaging mode using the UVW1 filter. We processed the OM data with the \texttt{omichain} SAS task and filtered the source list to retain only calibrated photometry of the target.
The average UVW1 magnitude for the whole 2016 campaign is $m_{\mathrm{UVW1}} = 15.36$ with a standard deviation of $0.03$ mag. The peak-to-peak variability amplitude is $\sim 0.17$ mag, hence the source has a stable UV emission varying by only $< 2\%$ during the whole XMM-Newton campaign.

\subsection{X-ray Spectral Analysis: Time and Flux-resolved Approaches}
To investigate spectral variability with high statistical significance, we adopted a strategy based on equal-count selections in both time and flux domains.
For the time-resolved analysis, we split the background-subtracted 7--10~keV light curve into five intervals, each containing an equal number of counts. This approach ensures that each time-resolved spectrum has comparable signal-to-noise in the iron K$\alpha$ region, optimizing our ability to study the Fe~XXV/XXVI features at high significance.
For the flux-resolved analysis, we divided the 3--10~keV light curve into five bands, again using thresholds that yield five spectra with equal total counts. This choice allows us to explore flux-dependent spectral changes across the broader continuum, minimizing potential biases from absorption features that may affect the higher energy band.
We also tested finer selections, including a ten-flux division following \citet{pinto_ultrafast_2018} and a 20-interval time-resolved selection. While these approaches are consistent with our main results, the lower signal-to-noise in each spectrum limits the robustness of the conclusions. Therefore, we focus on the five time and five flux selections, which provide a balance between temporal/flux resolution and statistical quality, enabling stronger constraints on spectral variability.
To mitigate the impact of statistical noise while preserving sensitivity to narrow spectral features, the spectra were rebinned using the \texttt{specgroup} task within the \texttt{SAS} software package. The binning was performed such that no spectral channel oversamples the instrumental energy resolution by more than a factor of three.
In our baseline analysis, we adopt a minimum of 5 counts per bin, a choice that provides improved sampling of the Fe~K$\alpha$ region and the associated outflow signatures, which would otherwise be excessively smoothed by more aggressive binning. Spectral fitting is carried out using the $W$ statistic in \texttt{XSPEC}, which is appropriate for Poisson-distributed data and remains valid across the full range of count regimes explored here.
We verified robustness by testing unbinned spectra and groupings of 5 and 25 counts/bin (maintaining bins $\geq$1/3 resolution element), and comparing $W$ and $\chi^2$ statistics. Results remain consistent within uncertainties. We adopt 5-count binning with resolution-based grouping and $W$ statistic, optimally balancing statistical robustness and spectral resolution.

\subsection{NuSTAR Data}
NuSTAR observed IRAS 13224-3809 for a total of $\sim$600~ks (July--August 2016). The source is very soft and NuSTAR has a smaller effective area than EPIC-pn, so the signal-to-noise ratio above $\sim$10~keV is intrinsically low. To optimize the signal-to-noise ratio in the 7--15 keV band we tested source and background apertures ($r_a = 40,35,30,25,20$ arcsec) and found a 25\arcsec\ source radius (50\arcsec\ background on the same chip) gave the best results; this radius encloses only an encircled energy fraction of $\sim$40\% \citep{nustar_psf_2014}. The ancillary response file produced by \texttt{nuproducts} includes an encircled energy fraction correction, so the aperture-dependent effective area loss is accounted for in our modeling. Only by stacking all NuSTAR exposures could we reliably recover the source up to $\sim$20 keV; for the average spectrum we applied coarse binning (8 channels per bin) to ensure $\gtrsim$25 counts/bin. Given the source variability (factor $\sim$20), the time-averaged NuSTAR spectrum is not suitable for flux-resolved fits, so NuSTAR is excluded from interval-resolved analyses and retained only for the stacked average spectrum (FPM A and B fitted separately and plotted as their average; Figure~\ref{fig:averagespec}).

\begin{figure}
  \centering
  \includegraphics{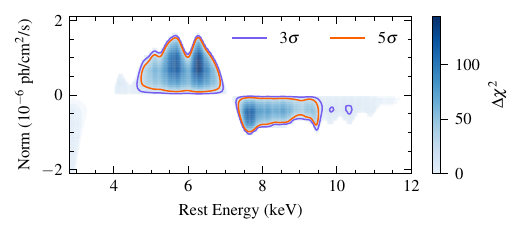}
  \caption{Result of a blind Gaussian line search performed on the average XMM EPIC-pn and NuSTAR FPM spectra, plotted in the 3-12 keV range in the IRAS 132224-3809 rest frame. The line width is fixed to $\sigma_{\rm{line}}=100$ eV. The ordinate axis shows the line normalization. Negative values indicate absorption lines. The purple and orange solid contours represent the 3$\sigma$ and 5$\sigma$ confidence levels corresponding to $\Delta \chi^2 = 11.8$ and $\Delta \chi^2 = 28.7$ for the addition of two degrees of freedom, respectively.}
  \label{fig:averagescan}
\end{figure}
\section{X-ray Average Spectrum}
\label{sec:average}

\begin{figure}
    \centering
    \includegraphics{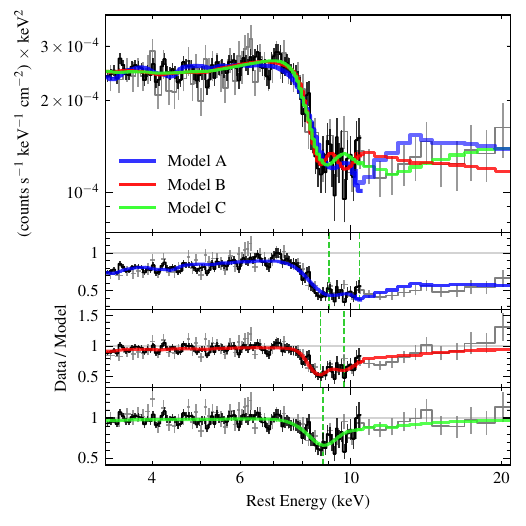}
    \caption{Average XMM-Newton EPIC-pn (black) and NuSTAR FPM (gray) spectra of IRAS 13224-3809 fitted with Model A (absorption only), Model B (absorption + Gaussian emission), and Model C (relativistic reflection) in the 3-20 keV range. 
    The three panels show the data-to-model ratio for each model, where we removed the absorption components to better visualize their effect on the continuum. The vertical dashed lines indicate the Fe K absorption centroids of each layer. Model A prefers lower-ionization absorbers to account for the curvature observed around 7 keV.
    See Table~\ref{tab:average_model_comparison} for details.}
    \label{fig:averagespec}
\end{figure}

We first analyzed the average spectrum from stacked XMM-Newton EPIC-pn and NuSTAR FPM observations (0.3--20 keV), focusing on $E > 3$ keV where the prominent soft excess is less dominant, enabling clearer assessment of continuum and absorption features. A simple absorbed power-law model ($\Gamma = 2.56 \pm 0.02$) provides a poor fit to the data, with reduced $\chi^2_\nu \sim 4.4$, indicating significant spectral complexity. To systematically identify additional features, we performed a blind Gaussian line scan (Appendix~\ref{sec:gauss_scan}), which revealed both broad emission and absorption structures across the spectrum (Fig.~\ref{fig:averagescan}).
We tested three distinct spectral models, building in complexity to explore the origin of these features. First, we introduced photo-ionized absorbers using \texttt{XSTARv2.58} table models (\citet{kallmann_xstar_1999}; see Appendix~\ref{sec:xstar} for further details). The photo-ionized gas in the \texttt{XSTAR} simulations is assumed to be optically thin and to have a single velocity component. One absorber layer improved the fit ($\chi^2_\nu = 1.53$); residuals in the Fe K region motivated a second component, yielding $\chi^2_\nu = 1.2$. This absorption-only model (Model A) is:
\begin{equation*}
\texttt{TBabs $\times$ XSTAR$_1$ $\times$ XSTAR$_2$ $\times$ zpowerlaw}.
\end{equation*}
Second, we added a broad Gaussian emission line to account for the 6--7 keV excess, yielding $\chi^2_\nu = 0.99$ ($\Delta\chi^2 = 46.5$, 3 d.o.f.) with an equivalent width (EW) of $1.42^{+0.40}_{-0.38}$ keV (Model B):
\begin{equation*}
\texttt{TBabs $\times$ XSTAR$_1$ $\times$ XSTAR$_2$ $\times$ (zpowerlaw + zgauss)}.
\end{equation*}
Finally, the best overall description was achieved using \texttt{relxill v2.6} \citep{dauser_relxill_2020}, which self-consistently computes intrinsic power-law continuum and relativistic reflection from an ionized Kerr disk. The \texttt{relxill} fit adopts a radial emissivity profile with index $\alpha = 3$, as expected for a co-planar point source in flat spacetime, and an accretion disk extending from the innermost stable circular orbit to $R_{\rm out} = 400\,R_{\rm g}$. We fixed $E_{\rm{cut}} = 250$ keV (typical and unconstrained by NuSTAR). A single absorber layer sufficed ($\Delta\chi^2 = 56.5$, 4 d.o.f.; $\chi^2_\nu = 0.95$), yielding (Model C):
\begin{equation*}
\texttt{TBabs $\times$ XSTAR $\times$ relxill}.
\end{equation*}
The model indicates reflection dominance, high iron abundance ($A_{\rm{Fe}} \sim 3$), and a highly ionized inner disk ($\log\xi > 3.5$ erg cm s$^{-1}$). We also tested whether a low cutoff ($<$100--200 keV) could mimic the spectral downturn, finding no improvement. Best-fit parameters are listed in Table~\ref{tab:average_model_comparison}; fitted spectra appear in Fig.~\ref{fig:averagespec}.

\begin{table}[ht]
\centering
\tiny
\begin{tabular}{l C{1.7cm} C{1.7cm} C{1.7cm}}
\toprule
\textbf{Parameter} & \textbf{Model A} & \textbf{Model B} & \textbf{Model C} \\
\midrule
\multicolumn{4}{c}{\textbf{Continuum}} \\
$\Gamma$ & $2.71^{+0.06}_{-0.07}$ & $2.44^{+0.04}_{-0.04}$ & $2.11^{+0.03}_{-0.05}$ \\
\texttt{norm} ($10^{-5}$) & $120 \pm 20$ & $41 \pm 3$ & $0.0054^{+0.0050}_{-0.0012}$ \\
\midrule
\multicolumn{4}{c}{\textbf{Broad Gaussian Emission}} \\
$E$ (keV) & $-$ & $6.2\pm0.3$ & $-$ \\
$\sigma$ (keV) & $-$ & $1.9^{+0.3}_{-0.2}$ & $-$ \\
EW (keV) & $-$ & $1.4\pm0.4$ & $-$ \\
\texttt{norm} ($10^{-6}$) & $-$ & $9.0^{+1.8}_{-1.4}$ & $-$ \\
\midrule
\multicolumn{4}{c}{\textbf{Relativistic Reflection}} \\
$a$ & $-$ & $-$ & $0.3^{+0.7}_{-0.3}$ \\
$i$ (deg) & $-$ & $-$ & $56^{+3}_{-2}$ \\
$\log\xi$ & $-$ & $-$ & $3.8\pm0.1$ \\
$A_{\rm{Fe}}$ & $-$ & $-$ & $3.0^{+1.0}_{-0.8}$ \\
$R_{\rm refl}$ & $-$ & $-$ & $>46$ \\
\midrule
\multicolumn{4}{c}{\textbf{Ionized Absorber 1}} \\
$N_{\rm H}$ ($10^{23}$ cm$^{-2}$) & $7.5^{+1.0}_{-0.7}$ & $7.9^{+12}_{-4.0}$ & $>15$ \\
$\log\xi$ (erg cm/s) & $3.2\pm0.04$ & $3.6^{+0.4}_{-0.2}$ & $4.5^{+0.1}_{-0.9}$ \\
$v_{\rm turb}$ (km/s) & $>25400$ & $14000^{+600}_{-400}$ & $>28100$ \\
$v_{\rm out}$ ($c$) & $0.32\pm0.01$ & $0.26\pm0.01$ & $0.27\pm0.01$ \\
\midrule
\multicolumn{4}{c}{\textbf{Ionized Absorber 2}} \\
$N_{\rm H}$ ($10^{23}$ cm$^{-2}$) & $8.9^{+1.3}_{-1.4}$ & $>171$ & $-$ \\
$\log\xi$ (erg cm/s) & $3.1\pm0.1$ & $4.5^{+0.1}_{-0.8}$ & $-$ \\
$v_{\rm turb}$ (km/s) & $>28400$ & $>24700$ & $-$ \\
$v_{\rm out}$ ($c$) & $0.48\pm0.01$ & $0.34 \pm 0.02$ & $-$ \\
\midrule
$\chi^2$/d.o.f &  253.85/212 &  207.29/209 &  204.71/211 \\
\bottomrule
\end{tabular}
\vspace{1em}
\caption{Comparison of best-fit parameters for Models A, B, and C. Missing parameters are indicated by $-$. Model A and B use two photo-ionized outflowing components, Model C needs just one ultra-fast outflow and includes a relxill reflection component. For Model C, continuum parameters ($\Gamma$, norm) refer to the intrinsic power law, while reflection parameters describe the reprocessed component; both are computed self-consistently within \texttt{relxill}.}
\label{tab:average_model_comparison}
\end{table}

\section{Flux- and Time-Resolved Spectral Analysis}
\label{sec:LMF_specs}

To investigate spectral variability, we applied the same three models tested on the average spectrum to all flux- and time-resolved intervals.
As a first step, we performed a blind Gaussian line scan across the spectra to identify significant emission and absorption features, guiding our choice of model components (Figures~\ref{fig:fluxscans} and \ref{fig:timescans}). We also report results from a purely phenomenological model comprising an absorbed power-law continuum plus Gaussian absorption lines with free line widths (Tables~\ref{tab:flux_phenom_comparison} and \ref{tab:time_phenom_comparison}). Unlike \citet{parker_response_2017} and \citet{chartas_variable_2018}, we left the line width free to vary to monitor possible changes in UFO broadening. We systematically find broadening values larger than the $\sigma=0.1$ keV assumed in previous works, which inflates the equivalent width (EW) measurements.

For model selection, we used the Akaike Information Criterion \citep[AIC;][]{burnham_anderson_2002, liddle_aic_2007}. To demonstrate the underlying physical consistency across models, we report the best-fit parameters and statistical improvement of the primary absorber across all intervals. Additional UFO components (e.g., the second absorber in Model A) were retained only when supported by strong statistical evidence, requiring $\Delta\mathrm{AIC} > 10$ (evidence ratio $P_0 = \exp(-\Delta\mathrm{AIC}/2) < 0.006$).

For Model A (absorption only), we began with a simple absorbed power-law continuum and iteratively added photo-ionized absorber components using \texttt{XSTAR} table models. Up to two absorbers were required in lower flux intervals, while a single absorber sufficed at higher fluxes (Tables~\ref{tab:flux_modelA_comparison} and \ref{tab:time_modelA_comparison}).

For Models B and C, we tested the addition of either a broad Gaussian emission line or a relativistic reflection component to the absorbed power-law continuum. For Model B, we allowed the Gaussian centroid, width, and normalization to vary but imposed a limit of $\sigma \le 1.5$ keV to maintain fit stability, as broader widths led to unphysical solutions and strong degeneracies with the continuum. This broad Gaussian is intended as a phenomenological proxy for wind emission: \citet{parker_degeneracy_2022} have shown that the 3--10\,keV emission feature in IRAS\,13224-3809 can be reproduced not only by relativistic reflection but also by P-Cygni profiles from a multi-component wind. We verified that varying the upper bound over the range $\sigma = 1.0$--$1.5$\,keV does not significantly alter the inferred UFO parameters (velocity, ionization, column density). For the relativistic reflection model, we fixed the spin parameter to $a=0.3$, the disk inclination to $i=56$ degrees, and the iron abundance to $A_{\rm{Fe}}=3$ as found in the average spectrum fit (\ref{tab:average_model_comparison}), to reduce parameter degeneracy and improve fit stability since we do not expect these parameters to vary on the timescales of two months. However, the spin parameter is unconstrained in the average spectrum fit, with the error range spanning from 0 to 1; $a=0.3$ is simply the face value from the fit, and any choice of spin within this interval does not strongly affect the results.
Within $1\sigma$, these parameters are consistent with those reported by \citet{parker_response_2017} ($a=0.989\pm0.001$, $i=58\pm1^\circ$, $A_{\rm{Fe}}=3.5\pm0.2$).
In higher flux intervals for Models B and C, spectra were well described by continuum and emission alone. In lower flux intervals, one additional UFO component is still required (see Tables \ref{tab:flux_modelB_comparison}, \ref{tab:time_modelB_comparison}, \ref{tab:flux_modelC_comparison}, \ref{tab:time_modelC_comparison}).
The absence of further absorbers is reasonable, as the spectral drop above 8 keV is already accounted for by the emission line blue wing or the reflection component.

\begin{figure}
    \centering
    \includegraphics{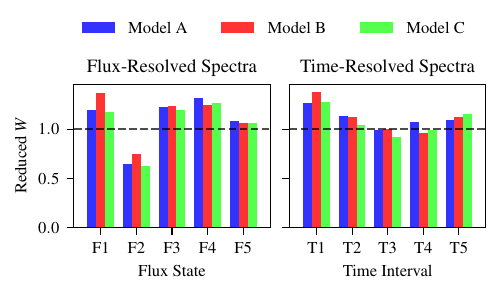}
    \caption{Histograms showing the distribution of reduced $W$ statistic ($W_\nu$) values for Models A, B, and C across all flux- and time-resolved spectra. Each model demonstrates comparable fit quality, with no single model consistently outperforming the others across the different intervals. This indicates that all three modeling approaches are viable for describing the spectral variability of IRAS 13224-3809 within the statistical uncertainties of the data.}
    \label{fig:fitquality}
\end{figure}

\begin{figure}
    \centering
    \includegraphics{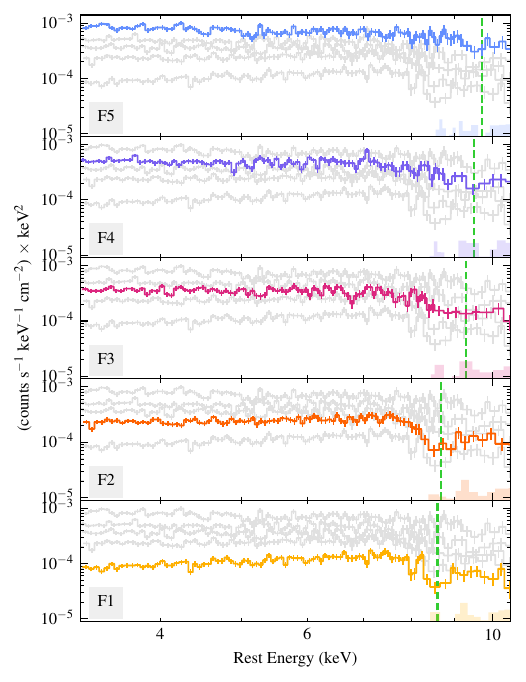}
    \caption{Best-fit models for the flux-resolved XMM-Newton EPIC-pn spectra of IRAS 13224-3809 in the 3-11 keV range. Data points are rebinned at 3$\sigma$ for easier visualization. Each panel corresponds to a different flux interval, labeled F1 through F5, with F1 being the lowest flux state and F5 the highest. The vertical dashed lines highlight the evolution of the centroid energy of the primary absorption feature associated with the UFO as phenomenologically modeled with a Gaussian.}
    \label{fig:fluxfits}
\end{figure}

\begin{figure}
    \centering
    \includegraphics{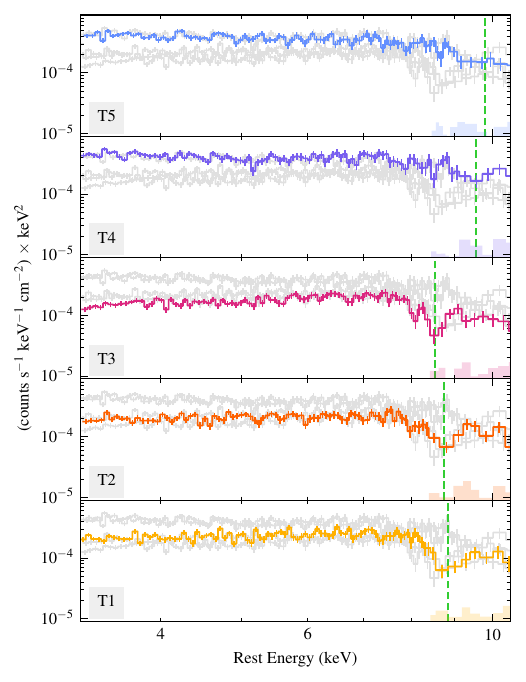}
    \caption{Same as Fig. \ref{fig:fluxfits} but for the time-resolved spectra labeled T1 through T5, with T1 being the earliest time interval and T5 the latest.}
    \label{fig:timefits}
\end{figure}

No single model was statistically preferred across all intervals, with fit statistics remaining comparable (Figure~\ref{fig:fitquality}). The best-fit models for flux- and time-resolved spectra are shown in Figures~\ref{fig:fluxfits} and \ref{fig:timefits}, demonstrating that all three approaches successfully describe spectral variability across different intervals.

All the column densities reported in this work for the UFO components are corrected for special relativistic effects following the prescription in \citet{luminari_wine_2024} (Eq. 14):
\begin{equation*}
N_H = \frac{1+v/c}{1-v/c} \, N_H^{\rm obs} \, .
\label{eq:abs-rel}
\end{equation*}
This relation allows us to derive the intrinsic $N_H$ from the observed value under the assumption of a radial, uniform outflow along the line of sight, with a single velocity per component, optically thin absorption, and no transverse velocity components or relativistic ray-tracing effects.

\subsection{Model Comparison}

\begin{figure*}
    \centering
    \includegraphics[width=\textwidth]{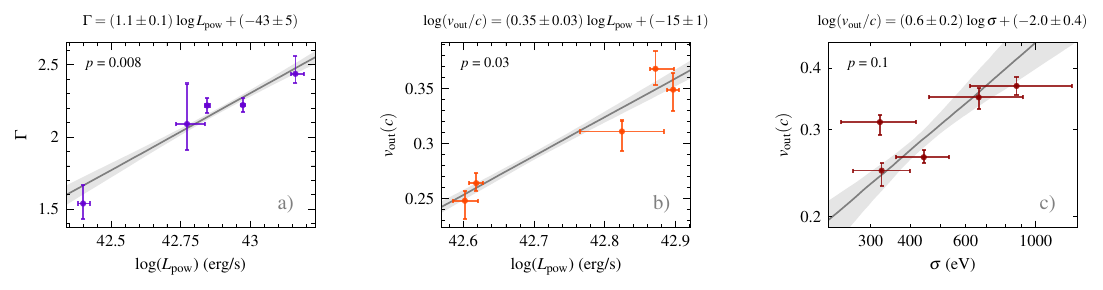}
    \caption{Correlations among key spectral and wind properties of IRAS 13224-3809. Left: Photon index $\Gamma$ versus 3–10 keV luminosity for the flux-resolved spectra (F1–F5), illustrating the \textit{softer-when-brighter} trend. Center: Outflow velocity of the ultra-fast outflow as a function of 3–10 keV luminosity for the time-resolved spectra (T1–T5), highlighting their coupled evolution. Right: Relation between UFO outflow velocity and Fe K$\alpha$ line width, tentatively indicating that faster winds are associated with greater velocity dispersion. Shaded regions represent 1$\sigma$ uncertainties on the best-fit correlations, derived from 10,000 bootstrap regressions sampling the error boxes. Each panel includes the $p$-value from an exact permutation test assessing null hypotheses of no correlation and the best fit relation.
    }
    \label{fig:correlations}
\end{figure*}

The results from the different modeling approaches (Models A, B, and C) provide complementary insights, although they also highlight some discrepancies and challenges in interpretation.
Figure \ref{fig:FLUX_TIME_recap} summarizes the evolution of the best-fit parameters derived from the spectral analysis of all flux- and time-resolved spectra. We adopted the three different models and a purely phenomenological modeling (power law plus Gaussian absorption lines only). In this latter model, we derive outflow velocities from the line centroid energy, assuming it associates with Fe XXVI K$\alpha$, as most of our best-fit photo-ionization modeling indicates (see Appendix \ref{sec:xstar}).
Model A (absorption only) consistently requires an absorber with high significance ($\gtrsim 3\sigma$) across all time- and flux-resolved intervals, yielding clear trends in the UFO parameters.
While the statistical significance of the UFO in Models B and C can decrease at higher fluxes where the emission components account for more of the spectral curvature, the inferred physical parameters and specifically the outflow velocities, remain remarkably consistent with those of Model A (see Appendix Fig.~\ref{fig:FLUX_TIME_recap}). This indicates that the detection of the wind is not an artifact of the continuum choice, but a robust feature whose significance is simply modulated by the underlying modeling of the Fe K emission. In flux-resolved spectra, although Models B and C do not strictly require an absorber in the highest flux bins ($\Delta\mathrm{AIC} < 6$), the inclusion of a UFO component in those fits yields velocity estimates that are fully compatible with the clear acceleration trend seen in Model A.
Importantly, we construct all time- and flux-resolved intervals to have comparable signal-to-noise ratios.
This ensures that variations in absorber detection reflect genuine spectral variability rather than statistical fluctuations.
Finally, all models recover the softer-when-brighter behavior of the continuum. In Model A, this trend is unambiguous ($\Gamma$ increasing from $\sim$1.5 to $\sim$2.5 with flux; Fig.~\ref{fig:correlations}, panel A), while Models B and C show a flatter trend, as the emission component favors steeper power-law slopes (Fig.~\ref{fig:FLUX_TIME_recap}, panel a, b). Due to the small sample size, we assess significance using an exact permutation test, which enumerates all $5! = 120$ rank orderings and requires no distributional assumptions. Model A yields $p = 0.008$ (the minimum attainable value for $n=5$), while Models B and C give $p = 0.06$ and $p = 0.025$, respectively. To test whether the relation remains non-flat when parameter uncertainties are included, we performed a bootstrap linear regression. Following \citet{laurenti_lines_2025}, we considered uncertainties in both the $x$ and $y$ directions. Recomputing the slope over 10,000 iterations, with each point drawn from a uniform distribution within its $1\sigma$ bounds, gives a positive slope in more than 99.9\% of realizations, strongly disfavoring a flat trend. This behavior likely reflects intrinsic changes in the corona or accretion flow, with higher flux states corresponding to a softer, more efficient Comptonization spectrum \citep[e.g.,][]{soboloewska_longterm_2009, serafinelli_serendipitous_2017}.

Regarding the specific parameters of the emission components, in Model C, the fits find an iron abundance $A_{\rm Fe} \sim 3$ times solar and a high reflection fraction, often exceeding 30. The spin parameter remains unconstrained ($0 < a < 1$), so we fixed $a=0.3$ from the average fit to reduce degeneracies. No significant trend with flux or time is evident for the disk ionization, which remains high ($\log\xi > 3.5$ erg cm s$^{-1}$) across all intervals. For Model B, the broad Gaussian emission line requires full widths at half maximum (FWHM) exceeding 2 keV, with the equivalent width remaining high across flux states and the centroid energy showing a modest luminosity dependence.


\section{Discussion}
\label{sec:discussion}

\subsection{Physical Plausibility of the Models}

The physical plausibility of the emission models remains questionable. Both the phenomenological broad Gaussian and the relativistic reflection model require extreme parameters, such as very high equivalent widths for the emission lines and high reflection fractions.
Our modeling confirms previous claims in the literature that relativistic reflection can reproduce the observed spectra. However, we find this interpretation incompatible with the clear UFO parameter trends reported elsewhere \citep[e.g.,][]{parker_response_2017,pinto_ultrafast_2018,jiang_15_2018}. Indeed, both flux- and time-resolved analyses show that when we include relativistic reflection (Model C), we detect the UFO only in the lowest flux or earliest time intervals. At higher fluxes or later times, we can adequately model the spectra without an absorber. Therefore, we cannot simultaneously claim a reflection-dominated spectrum and maintain the clear UFO parameter trends.

Previous studies investigating wind evolution have typically not applied a fully systematic relativistic reflection model. Reflection was usually tested only on selected spectra, while the broader wind analysis relied on phenomenological or photo-ionized absorption models. For example, \citet{pinto_ultrafast_2018} modeled the continuum as a power law plus a Gaussian convolved with a relativistic profile and a photo-ionized absorber. \citet{chartas_variable_2018} used a simple absorbed power law with Gaussian lines for the systematic UFO analysis. Similarly, \citet{midooka_radiatively_2023} used a power law with up to three Voigt absorption lines.
We further note that our spectral fitting covers only $E > 3$ keV. Relativistic reflection imprints diagnostic features across the full X-ray band. Restricting the bandpass reduces sensitivity to disk spin, inner emissivity, and ionization profile, making these parameters harder to constrain and the reflection scenario harder to test conclusively. However, extending the bandpass would not straightforwardly improve constraints on the reflection geometry. Previous broadband analyses of IRAS 13224-3809 \citep[e.g.,][]{pinto_ultrafast_2018, jiang_15_2018, jiang_xmm-newton_2022} were unable to model the full spectrum with a single reflection component: they required either an additional phenomenological Gaussian or a second independent reflection component to account for the strong soft excess below 2 keV. This additional complexity introduces further degeneracies that limit the constraining power even with broader energy coverage.
Model C faces several physical challenges:
\begin{itemize}
    \item The high iron abundance ($A_{\rm Fe} \sim 3$) is difficult to justify for a local Seyfert galaxy like IRAS 13224-3809, particularly for iron alone rather than for all elements.
    \item The reflection fraction is often extremely high (greater than 30), which is hard to explain within standard reflection scenarios. To achieve such high reflection fractions, one would need to invoke extreme light-bending effects or highly anisotropic emission geometries, which are not typically expected, given also that polarimetric studies favor an extended corona rather than a very compact one (e.g., \citealt{Giannolli_ixpe_2023}). Notably, such high reflection fractions are physically inconsistent with the baseline emissivity index $\alpha=3$, as they would require much more compact illumination. We tested steeper emissivity profiles (up to $\alpha=9$), but found that this does not solve the parameter degeneracies, with the black hole spin remaining unconstrained and the reflection fraction still extremely high.
\end{itemize}
While previous studies reported tighter spin constraints \citep[e.g.,][]{parker_response_2017}, our systematic analysis across all flux states reveals that CCD data do not robustly determine this parameter ($0 < a < 1$). The shallow, intrinsically broad line profiles, limited spectral resolution, and the restriction of the modeling to energies $E>3$ keV all contribute to the lack of meaningful spin constraints.

\citet{jiang_xmm-newton_2022} modeled IRAS 13224-3809 using reflection models with enhanced disk densities, but the same physical challenges persist: absorption requires an iron abundance of $\sim$3 solar for iron alone, a second independent reflection component with a different ionization state is needed for the soft excess, and even with enhanced disk densities the spin is constrained to near-maximal values ($a>0.98$).

Our application of the \texttt{relxill} model thus shows that adopting a relativistic reflection scenario requires extreme parameters and precludes a coherent characterization of the UFO evolution, consistent with previous reports \citep{parker_response_2017,jiang_15_2018,jiang_xmm-newton_2022}.

Model B (including a broad Gaussian emission line) inherits some of the problems of Model C concerning the absorption features. In the flux-resolved spectra, the presence of strong emission erodes the need for absorbers, making it difficult to track their behavior. However, in contrast to Model C, time-resolved analysis always requires an outflowing component with physical parameters consistent with those derived from Model A.
As noted in Section \ref{sec:LMF_specs}, the FWHM values for the emission component exceed 2 keV. However, to maintain fit stability, we had to limit the line width to $\sigma \le 1.5$ keV, since broader widths resulted in strong degeneracies with the continuum, where the power-law normalization becomes negligible. This limit is still large enough to encompass a broad, structured emission component as expected from wind emission. \citet{parker_degeneracy_2022} already demonstrated on IRAS\,13224-3809 that the 3--10\,keV emission is consistent with P-Cygni profiles from a multi-component wind, where reproducing the feature using dedicated radiative-transfer simulations required even two wind components. We should therefore read the broad Gaussian in Model B as a phenomenological stand-in for this wind emission. The inferred UFO parameters are robust against this choice, as verified over the range $\sigma = 1.0$--$1.5$\,keV. A self-consistent treatment using the WINE wind model \citep{liminari_wine_2018, luminari_wine_2024} will be presented in a forthcoming paper.

Despite their nature, Models B and C confirm that the data are compatible with an emission structure that accompanies the absorption, as previously shown by \citet{parker_degeneracy_2022}.

\subsection{Trends for UFO Parameters: Flux and Time Dependence}

A key result of our time-resolved analysis is the direct measurement of UFO acceleration in response to X-ray flares. The wind rapidly accelerates during luminosity increases, reaching velocities up to $\sim$0.38$c$ (see top panel of Fig.~\ref{fig:TimeUFO_Combined}), providing the first direct observational constraint on the wind acceleration timescale and its coupling to the central engine.\\
Regarding the ionized outflowing absorber, Model A consistently shows that the outflow velocity increases with higher fluxes, indicating a direct and dynamic response of the UFO to variations in the central engine. This velocity–luminosity correlation is robustly confirmed in both flux- and time-resolved analyses, demonstrating it is intrinsic to the source and not an artifact of the selection procedure (see top panel of Fig.~\ref{fig:TimeUFO_Combined}, center panel of Fig.~\ref{fig:correlations}). 
Model B independently confirms the velocity–luminosity trend and the wind acceleration. It recovers a significant absorber ($\Delta\mathrm{AIC} > 6$) in all five time-resolved intervals and in the three lowest flux intervals (F1–F3), with outflow velocities comparable with the ones inferred from Model A (Tables~\ref{tab:time_modelB_comparison} and \ref{tab:flux_modelB_comparison}). Where Model C detects the absorber (T1–T3 and F1–F3), its inferred velocities agree with Models A and B within $3\sigma$, consistently tracing the same velocity increase with luminosity. In the highest flux states (T4, T5, F4, F5), the reflection component in Model C introduces curvature in the iron K-band that absorbs the absorption residuals, which is a limitation at high flux rather than a contradiction of the velocity–luminosity trend seen robustly in the other two models. Although the statistical significance of the UFO in Models B and C drops below the $\Delta\mathrm{AIC} = 6 \approx 2 \sigma$ threshold in some of these high flux intervals, the physical outflow velocities inferred from these models remain consistent with Model A within their statistical uncertainties. This mutual agreement across all three models (Fig.~\ref{fig:FLUX_TIME_recap}, panels c, d) directly mitigates concerns of model degeneracy, demonstrating that our main physical conclusions about the observation of wind acceleration and the velocity-luminosity relation, are robust regardless of the baseline continuum model. Previous studies \citep{pinto_ultrafast_2018,chartas_variable_2018} reported the velocity-luminosity correlation, and our analysis strengthens this evidence by confirming it simultaneously in both flux and time domains, across independent spectral prescriptions, and by directly measuring the wind acceleration. Applying the same exact permutation test described above, we obtain $p = 0.025$ for Model A and $p = 0.008$ for Model B in the time-resolved intervals, supporting the correlation across independent spectral prescriptions. Furthermore, bootstrap linear regression that includes the measured parameter uncertainties gives a positive slope in more than 99.9\% of realizations, yielding for Model A the relation $\log (v_{\rm out}/c) = (0.9 \pm 0.2) \log(L_{3-10}) - (39 \pm 9)$.

Regarding the anti-correlation between EW and 3--10 keV luminosity reported by \citet{parker_response_2017} and \citet{pinto_ultrafast_2018} but not confirmed by \citet{chartas_variable_2018}, we find no significant trend; the EW is consistent with being constant across all flux and time intervals.
This result is consistent with the line scans shown in Fig.~\ref{fig:fluxscans} and Fig.~\ref{fig:timescans}, where the line normalization is plotted as a function of line energy. In lower flux states, the absorption lines appear more tightly localized, whereas in higher flux states the line detection region becomes more extended. As a consequence, the EW remains approximately constant across the intervals, although the detection of the line is more robust in lower flux states.
Allowing the line width $\sigma$ to vary confirms that the UFO features are systematically broader than 0.1~keV, in agreement with the turbulent velocities inferred from the XSTAR modeling, which are often in excess of 10,000~km s$^{-1}$.
In the Appendix (Fig.~\ref{fig:EW-flux}), we directly demonstrate the absence of a correlation between EW and luminosity in the time-resolved spectra, and compare our results with the best-fit correlation reported by \citet{parker_response_2017}.
The limited spectral resolution of the EPIC-pn data and the large observed equivalent widths limit our ability to fully disentangle the UFO physical properties, particularly between the column density and ionization. However, this degeneracy is not absolute: the ionization parameter is reasonably constrained across intervals ($\log \xi \approx 3.3$--$4.4$), and for $\log \xi \gtrsim 3.6$ erg cm/s the 3--10 keV opacity is dominated by H-like iron (Fe~\textsc{xxvi}), which reduces ambiguities in line identification (see Appendix~\ref{sec:xstar}).

\begin{figure}
    \centering
    \includegraphics{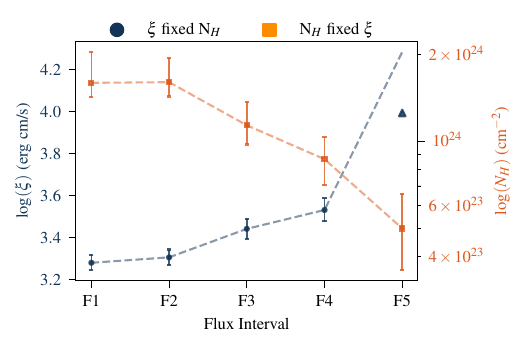}
    \caption{The ionization parameter ($\log\xi$) of the UFO and column density ($N_H$) corrected for special relativistic effects for the flux-resolved spectra (F1-F5) of IRAS 13224-3809. Here we are directly visualizing the trends when fixing one of the two parameters to its average value across the intervals.}
    \label{fig:xi-nh_fixed}
\end{figure}

In light of recent XRISM micro-calorimeter results (e.g., \citet{xrism_pdsNature_2025}), broad absorption troughs seen with CCDs are resolved into multiple narrow, kinematically distinct components. Therefore, our single-zone fits should be read as effective, ensemble descriptions: the measured $v_\mathrm{out}$ acts as a centroid-weighted velocity tracing the global wind kinematics, while the fitted $\log\xi$ and $N_H$ represent averaged phases of a structured, multiphase absorber rather than a unique homogeneous medium.
By contrast, $N_H$ measurements carry large uncertainties. If we fix one parameter to its average value (e.g., $\log \xi = 3.4$ erg cm/s or $\log N_H = 23.7$ cm$^{-2}$), the other shows a clearer trend with flux (Figure~\ref{fig:xi-nh_fixed}); when both are left free, the large statistical errors make the data consistent with a weak trend or with no trend. Several $N_H$ and $v_{\rm turb}$ estimates reach the boundaries of our model grids, as indicated by limits in the relevant tables, reducing the interpretability of trends for these variables, although the outflow velocity remains robustly constrained.
Importantly, $v_\mathrm{out}$ is robustly measured since it primarily sets the line centroid, and does not display the same degeneracy. If $N_\mathrm{H}$ and $\log \xi$ were fully degenerate, we would expect a correspondingly larger uncertainty in $v_\mathrm{out}$, which we do not observe.
Thus, while our CCD-based errors are large, our results are broadly consistent with previously reported trends (e.g., \citealt{pinto_ultrafast_2018}) but do not permit equally tight independent constraints on both $N_\mathrm{H}$ and $\log \xi$.
This complex behavior finds a qualitative physical interpretation within the framework of Chaotic Cold Accretion (CCA; e.g., \citealt{gaspari_cca_2013, gaspari_raining_2017}). In this scenario, the multiphase rain and cloud structure is inherently stochastic and variable, which naturally explains the (i) large scatter and non-monotonicity observed in $\xi$ and $N_H$, and (ii) the episodic appearance of the absorption features, as clouds condense and dissolve or pass transiently through the line of sight.
Multiphase wind structure is also predicted by complementary physical mechanisms. Radiation-hydrodynamic simulations show that super-Eddington flows fragment into clumpy structures via Rayleigh-Taylor instabilities on hour-to-day timescales \citep{takeuchi_clumpy_2013}, consistent with our observed variability. Similarly, thermal instability rapidly condenses initially smooth outflows into multiphase clumps when plasma cools below the equilibrium curve, with growth times comparable to our observed acceleration timescale \citep{dannen_clumpy_2020, waters_multiphase_2021}. These mechanisms naturally produce the clumpy substrate upon which magnetic acceleration operates, potentially explaining both the structured absorption features and the rapid velocity response to X-ray flares we observe.

\subsection{Degeneracies and Modeling Challenges}
\label{sec:degeneracies}

The application of Model~A sometimes results in the detection of a second UFO component ($v_{\rm out} \sim 0.4c$) in low-flux intervals (either flux- or time-selected). While multiple components are plausible (e.g., MCG-03-58-007 \citealt{braito_stratified_2021} or PDS 456; \citealt{xrism_pdsNature_2025}), this is model-dependent: in Models B and C, steeper continua allow Fe~K$\beta$ or emission wings to explain these features without additional absorbers. Our data cannot robustly constrain multiple components.
When considering the average spectrum, Model~A gives a poorer fit ($\chi^2_\nu \sim 1.2$) than Models B and C ($\chi^2_\nu \sim 0.95$--$0.99$), suggesting that the stacked spectrum contains genuine broad emission features near 6--7 keV that cannot be reproduced by absorption models alone. In Model~A, the fit compensates by lowering the ionization parameter ($\log\xi \sim 3.1$ erg cm/s in the average, versus $\sim3.6$ erg cm/s in the resolved intervals), enhancing lower-ionization transitions that broaden the continuum curvature. A similar approach was adopted by \citet{midooka_radiatively_2023}, though it does not adequately reproduce the emission-like feature near 6--7 keV.

To check whether stacking spectra with different absorption properties could artificially produce emission features, we simulated and stacked synthetic spectra using the observed trends in continuum slope and UFO velocity. The resulting stacked spectrum showed only smeared absorption with no significant emission, confirming that the emission seen in the average spectrum is intrinsic (see Figure~\ref{fig:stacked_simulation}). We cannot rule out a contribution from variability below 4 keV, where the soft spectrum is not well understood.

Model differences appear above 10 keV: Models~A and B predict a flatter spectrum; Model~C shows a Compton hump at 20--30 keV. However, NuSTAR lacks sufficient signal-to-noise above 10 keV, and the Swift Burst Alert Telescope is insensitive to the flux predicted by Model~A ($F_{\rm{14-195\,keV}} \sim 6 \times 10^{-13}$ erg s$^{-1}$ cm$^{-2}$, well below its sensitivity limit). Current hard X-ray data cannot discriminate.

In summary: Model~A (absorption only) consistently reveals clear trends in the UFO parameters across all intervals, but provides a poorer fit to the average spectrum and cannot reproduce the observed broad emission features. Model~B (absorption plus broad Gaussian emission) also tracks the UFO parameters well, but requires an extremely broad emission line (FWHM $\gtrsim$ 2~keV). Model~C (relativistic reflection) fits the average spectrum best, but requires extreme parameters such as very high iron abundance, reflection fractions, and broad emission lines, and in high-flux states, often does not require significant absorption, weakening constraints on UFO variability. We caution that these conclusions rest on fitting the $E > 3$ keV band alone; broadband coverage would provide stronger leverage on the reflection geometry and on the degeneracy between reflection and absorption. This highlights the need for more advanced models (e.g., those incorporating clumpy, multi-phase outflows or wind emission features) and higher-sensitivity broad-band data from future missions to robustly disentangle absorption and emission components in AGN spectra.

\subsection{Width of UFO absorption features and implications for UFO structure}
A persistent feature in our analysis, regardless of the specific model employed, is the significant width of the absorption features. When left free to vary, the turbulent velocity ($v_{\rm turb}$) consistently converges to values exceeding $10\,000$ km s$^{-1}$, and often hits the upper limit of our model grid ($30\,000$ km s$^{-1}$). This is consistent with previous studies of this source, which required similarly large broadening to fit the data.
In the context of recent high-resolution X-ray spectroscopy, such extreme broadening requires careful interpretation. Groundbreaking results from the XRISM satellite, starting with PDS 456 \citep{xrism_pdsNature_2025,noda_iras_2025,mizumoto_pg1211_2025}, have revealed that broad absorption troughs observed in CCD spectra may not be due to single, intrinsically broad lines. Instead, they are resolved into complex "forests" of narrower lines ($\sigma \sim 1000$ km s$^{-1}$) originating from multiple clumps with distinct line-of-sight velocities.
It is highly probable that the "broad" UFO we observe in IRAS 13224-3809 follows this pattern. The large effective width likely represents the velocity dispersion of a clumpy, multi-phase outflow rather than turbulence within a homogeneous gas. If we consider the velocity range implied by our best-fit widths, the absorption trough spans from approximately $0.2c$ to $0.4c$. This suggests we are observing a superposition of kinematic components ranging from slower, perhaps newly launched gas, to faster, fully accelerated clumps. This interpretation aligns with the rapid variability we observe: as the central source flares, we may be seeing the sudden acceleration of specific clumps or the ionization of new velocity components within this broad kinematic structure.
High-resolution micro-calorimeter spectroscopy with XRISM and NewAthena will be crucial for resolving the wind substructure.

We note a tentative positive correlation between the outflow velocity and the line width (see panel C in Fig.~\ref{fig:correlations}), where higher velocities correspond to broader features. Its exact permutation significance is low ($p = 0.1$), and therefore this trend should be treated with caution. A bootstrap linear regression that includes the measured uncertainties yields a positive slope in most realizations, with $\log (v_{\rm out}/c) = (0.5 \pm 0.2) \log \sigma_{\rm Fe} + (0.0 \pm 0.1)$. This is consistent with the idea that faster winds are more structured, with a larger velocity dispersion among the absorbing material.
A similar but even weaker trend links velocity dispersion and X-ray continuum luminosity.
A positive correlation between outflow velocity and line width, as hinted in our time-resolved spectra, is found in larger AGN samples. \citet{laurenti_lines_2025} recently report that UFOs with higher velocities tend to show broader absorption features, based on a systematic analysis of over 100 AGN outflows. This trend may indicate that faster winds are more structured or clumpy, with a larger velocity dispersion among the absorbing material. Our results for IRAS 13224-3809 are consistent with this broader picture, but given the limited number of data points and the low spectral resolution of EPIC-pn, this trend should be considered tentative.

\subsection{Bolometric Luminosity}
\label{sec:bolometric}
To estimate the total radiation output, we complement the X-ray luminosity with a UV measurement that is robust against variability. Using a typical X-ray luminosity ($\log L_{\rm X} \sim 42.8$ erg s$^{-1}$) and the population-averaged bolometric correction ($K_{\rm X} \sim 15$; \citealt{duras_bolometric_2020}), we obtain $\log(L_{\rm bol, X} / \text{erg s}^{-1}) \approx 44.1$. We then used the XMM-Newton OM UVW1 filter ($\lambda_{\rm eff} \approx 2910$ \AA) photometry \citep{vagnetti_variability_2010, laurenti_investigating_2024}.
We corrected the flux density for Galactic extinction \citep{fitzpatrick_correcting_1999} and then extrapolated the monochromatic luminosity to rest-frame 3000 \AA\ assuming a canonical UV spectral slope of $\alpha_{\nu} = -0.5$ ($f_\nu \propto \nu^{\alpha}$) \citep{richards_sed_2006}. The resulting isotropic luminosity ($L_{\rm iso}$) was calculated using the log-linear bolometric correction provided by \citet{runnoe_updating_2012}:
\begin{equation*}
    \log(L_{\rm iso} / \text{erg s}^{-1}) = 1.85 + 0.97 \log(\lambda L_{\lambda, 3000} / \text{erg s}^{-1}) \, .
\end{equation*}
Finally, we applied the recommended anisotropy correction factor of $0.75$ \citep{runnoe_updating_2012} to account for the geometric bias of the accretion disk viewing angle, yielding our final, sustained bolometric luminosity estimate of $\log(L_{\rm bol} / \text{erg s}^{-1}) \approx 44.38$, or $L_{\rm bol} \sim 2.4 \times 10^{44}$ erg $\text{s}^{-1}$. This value is adopted for all subsequent outflow momentum and power calculations. We note that, within the typical bolometric correction uncertainty of $\sim$0.37~dex \citep{duras_bolometric_2020}, the UV-based bolometric luminosity is consistent with the X-ray-derived estimate.

\subsection{UFO Acceleration and Energetics}
Using the average values from the time-resolved analysis with Model A, we estimate the energetics of the UFO in IRAS 13224-3809. For these calculations, we adopt a black hole mass of $M_{\rm BH} = 10^7\,M_\odot$ as estimated by \citet{emmanopulos_mass_2014} from X-ray reverberation measurements. Adopting typical parameters from our analysis ($v_{\rm out} \sim 0.3c$, $N_H \sim 10^{24}$ cm$^{-2}$, $\log\xi \sim 3.7$) and the relation in \citet{tombesi_unification_2013}, we compute the minimum launching radius assuming $v_{\rm out} = v_{\rm esc}$: 
\begin{equation*}
    r_{\rm min} = 2GM_{\rm BH}/v_{\rm out}^2 \sim 3.3 \times 10^{13} \text{ cm} \sim 20 r_g \, .
\end{equation*}
The mass outflow rate is then:
\begin{equation*}
\dot{M}_{\rm out} = 4\pi C_f \mu m_p r_{\rm wind} N_H v_{\rm out} \sim 0.42 \, C_f \, M_\odot/\text{yr}
\end{equation*}
where $C_f$ is the covering factor of the wind, $\mu \sim 1.4$ is the mean atomic mass per proton, and $m_p$ is the proton mass and we used the minimum radius $r_{\rm min}$ as an estimate for $r_{\rm wind}$. Since both $\dot{M}_{\rm out}$ and $\dot{E}_{\rm kin}$ are linear in $C_f$ and $r_{\rm wind}$, adopting $r_{\rm min}$ and a covering factor $C_f = 0.3$, as representative of UFO population studies \citep{tombesi_evidence_2011, matzeu_subways_2023}, yields robust estimates for the energetic quantities. With these choices, $\dot{M}_{\rm out} \sim 0.13\,M_\odot$/yr and a kinetic power $\dot{E}_{\rm kin} = \frac{1}{2} \dot{M}_{\rm out} v_{\rm out}^2 \sim 1.1 \times 10^{44}$ erg/s.
We note that in some time-evolving studies where the absorber distance can be measured \citep[e.g.,][]{serafinelli_time_2025}, the wind velocity falls below the local escape velocity, indicating a failed wind destined to fall back.
The time evolution of the UFO velocity in IRAS 13224-3809 also provides valuable insight into the acceleration mechanism at play. Using the time-resolved spectra, we estimate the wind acceleration by computing the velocity difference between consecutive time intervals divided by the time separation between their mean observing times. The outflow velocities are taken from Model A. Model B recovers the absorber with $\Delta\mathrm{AIC} > 6$ in all cases and the inferred outflow velocity is consistent with Model A within the uncertainties (Table~\ref{tab:time_modelB_comparison}). The acceleration measurement is therefore not contingent on a flat power law continuum assumption of Model A.

\begin{figure}
    \centering
    \includegraphics{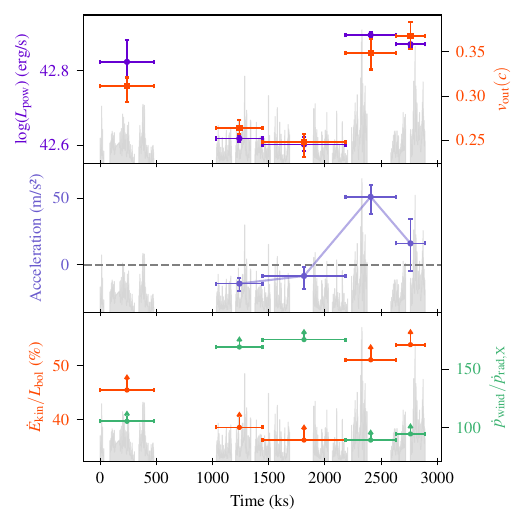}
    \caption{Three-panel summary of the time-resolved properties of the ultra-fast outflow (UFO) in IRAS 13224-3809, fitted with Model A (absorption only). Top: 3–10 keV power-law luminosity and UFO outflow velocity as a function of time, showing their coupled evolution. Middle: UFO acceleration, derived from velocity changes between intervals, highlighting rapid increases during bursts. Bottom: UFO kinetic power expressed as a percentage of the adopted bolometric luminosity; the right axis shows the wind momentum rate normalized to the X-ray radiation momentum rate. The left axis traces how energetic the wind is, while the right axis traces how difficult it is for the instantaneous X-ray radiation field alone to account for that wind momentum. In all panels, the background displays the 3–10 keV light curve for context.}
    \label{fig:TimeUFO_Combined}
\end{figure}

The outflow responds promptly to changes in the X-ray continuum, with a velocity increase of $\Delta v \approx 0.1c$ from interval T3 to T4 on a timescale of about 6 days ($\Delta t \approx 590$~ks), yielding an acceleration of $\sim50$~m~s$^{-2}$ (Fig.~\ref{fig:TimeUFO_Combined}, middle panel); Model B yields a lower estimate of $\sim30$~m~s$^{-2}$, still a significant increase. This rapid coupling to the X-ray luminosity suggests the wind is highly sensitive to the physical processes powering the corona.

To quantify the contribution of radiation pressure to the UFO acceleration, we estimated the acceleration imparted to a column of gas according to \citet{gu_ngc3783_2025}:
\begin{equation*}
a_{\rm rad} = \frac{\int F(E)\,[1-T(E)]\,dE}{c \, \mu \, m_p \, N_H},
\end{equation*}
where $F(E)$ is the incident X-ray flux at the wind location, $T(E)$ is the transmission as a function of energy, $c$ is the speed of light, $\mu$ is the mean atomic mass per proton, $m_p$ is the proton mass, and $N_H$ is the column density.
We consider the observed 3--10 keV X-ray luminosity and, using the average $N_H$ and $\xi$ values (accounting for their uncertainties), extract the multiplicative opacity spectrum from our XSTAR absorption model. By integrating this spectrum over the 3--10 keV band, we obtain the total band-averaged transmission, yielding $T = 0.89$. This value is representative of all time intervals given the uncertainties and degeneracy between $\xi$ and $N_H$ ($\log\xi \simeq 3.7$ erg cm/s, $\log N_{\rm H} \simeq 24$ cm$^{-2}$). The choice of energy range is not restrictive, as the wind imprints are confined to the 3--10 keV band, with negligible impact outside this interval. Moreover, while the X-ray luminosity shows strong variability, the bolometric luminosity, dominated by the optical emission, exhibits no significant changes on these timescales. This indicates that the UFO responds specifically to variations in the inner disk/corona rather than to global changes in the accretion disk luminosity.
We place the wind at the minimum launching radius $r_{\rm min}$, representing the most favorable configuration for radiation driving. Under these assumptions, we obtain $a_{\rm rad} \sim 10$~m~s$^{-2}$, which is a factor of $\sim3-5$ below the observed peak acceleration. This estimate should be regarded as an optimistic upper limit, as it assumes high column densities and the smallest possible wind radius. Even in this extreme scenario, radiation pressure alone cannot account for the prompt and strong acceleration of the UFO.
Instead, the immediate response and high acceleration favor magnetic processes, such as magnetic reconnection or magnetically driven winds, as the dominant mechanism, as recently proposed for NGC~3783 and NGC 4051 using XRISM Resolve observations \citep{gu_ngc3783_2025,reeves_ngc4051_2026}. In this scenario, reconnection events in the inner accretion disk and/or corona can rapidly release stored magnetic energy, simultaneously producing X-ray flares and accelerating ultra-fast outflows. The reconnection timescale is set by the Alfvén speed, which can approach relativistic values in AGN coronae, allowing for nearly instantaneous wind acceleration. This naturally explains the tight correlation between continuum flares and UFO velocity variations observed in IRAS~13224-3809, in analogy with solar coronal mass ejections. Our results therefore favor a magnetic origin for the UFO acceleration.

To assess the energetics and feedback potential of the ultra-fast outflow (UFO) in IRAS~13224$-$3809, we computed its momentum rate and kinetic power using our estimates. To minimize the impact of parameter uncertainties, we again adopted representative average values for the ionization parameter and column density, which dominate the systematic uncertainties in the energetics.
The UFO momentum rate and kinetic power were computed as $\dot{p}_{\rm wind}=\dot{M}_{\rm out}v_{\rm out}$ and $\dot{E}_{\rm kin}=\tfrac{1}{2}\dot{M}_{\rm out}v_{\rm out}^2$, respectively.
Adopting a covering factor $C_f=0.3$ and assuming the UFO is launched at the escape radius $r_{\rm min}$, the UFO momentum rate exceeds the radiation momentum rate $\dot{p}_{\rm rad}=L_X/c$, with $\dot{p}_{\rm wind}^{\rm min}\gtrsim 60\,\dot{p}_{\rm rad}$ (bottom panel of Fig.~\ref{fig:TimeUFO_Combined}). Under these assumptions, this momentum excess disfavors single-scattering radiation pressure as the dominant driving mechanism and instead points to either highly efficient multiple scattering or an additional driving agent, such as magnetic stresses.
The detailed trends in the bottom panel are nevertheless model dependent, since they rely on the conservative choice $R=r_{\rm min}$, which should be regarded as providing lower-bound energetics rather than a unique wind geometry. For the momentum ratio we use the 3--10 keV X-ray luminosity to compute $\dot{p}_{\rm rad}=L_X/c$, since the UV-based bolometric luminosity remains constant throughout the campaign and cannot track the observed variability. The UFO responds directly to changes in the X-ray continuum, so this choice allows us to follow the evolution of $\dot{p}_{\rm rad, X}$ and compare it meaningfully to the UFO momentum rate. By contrast, the kinetic-power fraction in the bottom-left panel is normalized by the constant UV-based bolometric luminosity adopted in Section~\ref{sec:bolometric}. Using the constant UV-based bolometric luminosity would yield only a single, static value $\dot{p}_{\rm wind} \simeq 3\,\dot{p}_{\rm rad,UV}$; even then, the UFO momentum rate is significantly larger than the radiation momentum rate.
The kinetic power of the outflow reaches $\dot{E}_{\rm kin}\sim 1.1 \times 10^{44}$~erg~s$^{-1}$, corresponding to $\sim 45\%$ of the bolometric luminosity, as shown in the bottom panel of Fig.~\ref{fig:TimeUFO_Combined}. This value comfortably exceeds the canonical $\sim0.5$--$5\%\,L_{\rm bol}$ threshold required for effective AGN feedback \citep{king_powerful_2015}, implying that the UFO in IRAS~13224$-$3809 is capable of exerting substantial influence on its host galaxy \citep{gaspari_windrev_2020}.


\section{Conclusions}
\label{sec:conclusions}

We presented a comprehensive time- and flux-resolved spectral analysis of the Narrow-Line Seyfert~1 galaxy IRAS~13224--3809, based on the 2016 \textit{XMM-Newton} (1.5~Ms) and \textit{NuSTAR} (500~ks) monitoring campaign. By systematically exploring three spectral frameworks (photo-ionized absorption, Model~A; phenomenological emission, Model~B; and relativistic reflection, Model~C), we draw the following main conclusions.

\begin{enumerate}

\vspace{0.5em}
\item Model comparison and physical interpretation.
All three spectral models provide statistically comparable fits and independently confirm the presence of a highly ionized, variable UFO. However, they differ substantially in physical plausibility.
Model~A (absorption-dominated) yields the most coherent and physically interpretable trends for the wind properties across both flux- and time-resolved spectra.
Model B tracks wind parameters with similar robustness. 
The inclusion of a broad Gaussian emission component improves the fit to the average spectrum, serving as a phenomenological proxy for UFO emission. 
However, in flux- and time-resolved spectra, this component requires an extreme width that becomes degenerate with the continuum slope. 
This degeneracy challenges the physical interpretation of the model and suggests that a single Gaussian does not adequately capture the complex emission structure of the source.
In contrast, Model~C (relativistic reflection) requires extreme and challenging parameters (reflection fractions $R \gtrsim 30$ and iron abundances $A_{\rm Fe} \gtrsim 3\,A_{\rm Fe,\odot}$), and the dominance of reflection features suppresses the detectability of the UFO at high fluxes.

\vspace{0.5em}
\item Velocity--luminosity correlation.
We robustly confirm a positive correlation between the UFO velocity ($v_{\rm out}$) and the X-ray luminosity ($L_{\rm X}$). The outflow responds dynamically to changes in the source brightness, with velocities increasing from $\sim0.25c$ in low-flux states to $\sim0.37c$ during flaring episodes.

\vspace{0.5em}
\item Line width and equivalent width behavior.
No significant anti-correlation is found between the EW of the absorption features and the source flux; the EW remains consistent with being constant across all intervals.
The UFO absorption features are intrinsically broad ($\sigma \gtrsim 300$~eV), implying turbulent velocities of $v_{\rm turb} \gtrsim 10^{4}$~km~s$^{-1}$.
We find tentative evidence for a positive correlation between outflow velocity and line width, suggesting that faster winds are increasingly kinematically complex, possibly reflecting the velocity dispersion within a clumpy, multi-phase outflow consistent with Chaotic Cold Accretion (CCA) predictions.

\vspace{0.5em}
\item Acceleration mechanism.
The UFO reacts rapidly to X-ray flares, reaching accelerations of up to $\sim50$~m~s$^{-2}$. Radiation pressure alone cannot account for this behavior, as the inferred force exceeds the expected radiative force by a factor of $\sim5$.
This strongly favors a magnetic driving mechanism, potentially associated with magnetic reconnection events in the inner accretion disk, analogous to solar coronal mass ejections. Importantly, this conclusion is robust against the choice of spectral model used to infer the outflow velocity.

\vspace{0.5em}
\item Energetics and feedback implications.
Adopting a representative covering factor ($C_f=0.3$) and the minimum launching radius ($r=r_{\rm min}$), the kinetic luminosity reaches $\dot{E}_{\rm kin} \sim 10^{44}$~erg~s$^{-1}$, corresponding to nearly half of the bolometric luminosity.
The UFO momentum rate significantly exceeds the radiation momentum rate ($\dot{p}_{\rm wind} \gtrsim 60\,\dot{p}_{\rm rad}$), demonstrating that the UFO in IRAS~13224--3809 is energetically capable of driving substantial AGN feedback on its host galaxy.

\end{enumerate}

The combination of near-Eddington kinetic power, large momentum flux, and compact launch radius places this source among the most powerful known AGN winds, comparable to those observed in luminous quasars such as PDS~456 \citep[e.g.,][]{xrism_pdsNature_2025}.
Future work will aim to resolve the remaining spectral degeneracies by applying the \textsc{WINE} model, which self-consistently accounts for both emission and absorption from the outflow. This approach promises a more unified picture of the wind structure and dynamics.

\begin{acknowledgements}
FT, EP, and ML acknowledge funding from the European Union - Next Generation EU, PRIN/MUR 2022 (2022K9N5B4).
MG acknowledges support from the ERC Consolidator Grant \textit{BlackHoleWeather} (101086804). RM acknowledges financial support from the INAF Scientific Directorate.
RS acknowledges funding from the CAS-ANID grant No. CAS220016.
LZ acknowledges financial support from the Bando Ricerca Fondamentale INAF 2022 Large Grant “Toward an holistic view of the Titans: multi-band observations of z > 6 QSOs powered by greedy supermassive black holes” and from the European Union - Next Generation EU, PRIN/MUR 2022 2022TKPB2P - BIG-z.
PC thanks Flor Arevalo Gonzalez and Vicente Madurga Favieres for useful discussions. AI tools (Gemini and Claude) were employed for language refinement and technical code assistance. The integrity of the research, including all analysis and results, was ensured through independent verification by the authors.
Figures in this work were produced using the \texttt{Pubplotlib} Python package.
\end{acknowledgements}

\bibliographystyle{aa}
\bibliography{iras13224}

\appendix

\section{Observations and Data Availability}
\begin{figure}[h!]
    \centering
    \includegraphics{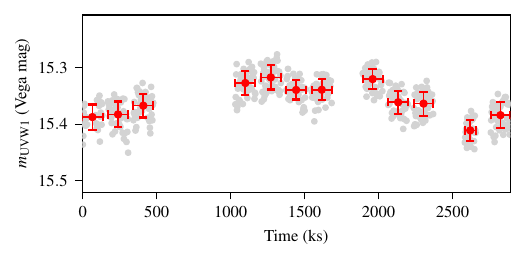}
    \caption{XMM-Newton Optical Monitor (OM) UVW1 light curve of IRAS 13224-3809 during the 2016 campaign. The source shows a stable UV emission with a peak-to-peak variability amplitude of only $\sim 0.17$ mag ($<2\%$) over the entire observation period. The average magnitude is $m_{\mathrm{UVW1}} = 15.36$ with a standard deviation of $0.03$ mag.}
    \label{fig:OM_lightcurve}
\end{figure}

The data used in this work are publicly available archival data from the XMM-Newton and NuSTAR observatories. The list of observations used in this work is shown in Table \ref{tab:obs}. In figure \ref{fig:OM_lightcurve} we show the XMM-Newton Optical Monitor (OM) UVW1 light curve of IRAS 13224-3809 during the 2016 campaign.
All XMM-Newton EPIC-pn observations were performed in Large Window Mode.
\begin{table}[h!]
  \tiny
    \centering
    \begin{tabular}{c c c C{0.5in} C{0.5in}}
        \hline
        Date & Telescope & OBS ID & Net Exposure (ks) & Count Rate (counts/s) \\
        (1) & (2) & (3) & (4) & (5) \\
        \hline
        2016-07-08 & XMM & 0780560101 & 34 & 1.5 \\
        2016-07-10 & XMM & 0780561301 & 116 & 1.6 \\
        2016-07-12 & XMM & 0780561401 & 102 & 1.5 \\
        2016-07-20 & XMM & 0780561501 & 106 & 1.1 \\
        2016-07-22 & XMM & 0780561601 & 96 & 2.5 \\
        2016-07-24 & XMM & 0780561701 & 94 & 1.1 \\
        2016-07-26 & XMM & 0792180101 & 111 & 1.0 \\
        2016-07-30 & XMM & 0792180201 & 113 & 1.5 \\
        2016-08-01 & XMM & 0792180301 & 81 & 0.73 \\
        2016-08-03 & XMM & 0792180401 & 66 & 3.6 \\
        2016-08-07 & XMM & 0792180501 & 103 & 1.3 \\
        2016-08-09 & XMM & 0792180601 & 98 & 3.4 \\
        2016-07-08 & NuSTAR & 60202001002 & 73 & 0.013 \\
        2016-07-10 & NuSTAR & 60202001004 & 72 & 0.015 \\
        2016-07-12 & NuSTAR & 60202001006 & 72 & 0.016 \\
        2016-07-23 & NuSTAR & 60202001008 & 75 & 0.02 \\
        2016-07-27 & NuSTAR & 60202001010 & 67 & 0.013 \\
        2016-08-01 & NuSTAR & 60202001012 & 184 & 0.02 \\
        2016-08-08 & NuSTAR & 60202001014 & 146 & 0.026 \\
        \hline
    \end{tabular}
    \vspace{1em}
    \caption{List of XMM EPIC-pn and NuSTAR FPM observations used in this work. Column (1) refers to the start date of each exposure. Column (4) lists background-subtracted source counts in the 0.3-10 keV band for the XMM-Newton EPIC-pn instrument and in the 3-20 keV band NuSTAR FPM instruments while (5) contains the average background-subtracted count-rate in the observation. For NuSTAR, an average value between the FPMA and FPMB modules is indicated.}
    \label{tab:obs}
\end{table}

\section{Gaussian Line Scan and Significance Estimation}
\label{sec:gauss_scan}
To identify line features above the continuum across a broad energy range, we conducted a blind search using an absorbed power-law continuum model combined with a Gaussian line component, which could be either in emission or absorption.
Line energies greater than 3 keV were scanned in the source rest frame, with 10 eV increments. At each step, the normalization of the line was varied, and any resulting statistical improvement in the fit was noted.
The line width was fixed to $\sigma_{\rm{line}}=100$ eV since we know from previous analyses that we are dealing with broad features and from the fact we are hitting the instrumental resolution limit.
To draw the significance contours of the detected features, we used the likelihood ratio test. Alternatively, significance can be estimated using the F-test, as commonly done in previous UFO studies, or information criteria such as AIC; we tested all these approaches and all yielded consistent results. To remain test-independent, we generally report fit improvement.

\begin{figure}
    \centering
    \includegraphics[width=\linewidth]{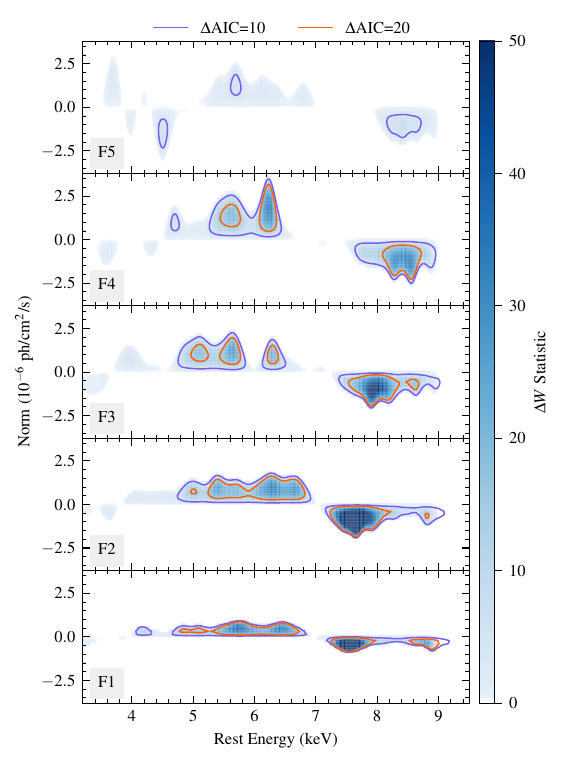}
    \caption{
        Gaussian line scans performed on the flux-resolved spectra of IRAS 13224–3809 in the 3–10 keV range in the source rest frame. The line width is fixed to $\sigma_{\rm{line}}=100$ eV, and the ordinate axis shows the best-fitting line normalization, with negative values indicating absorption features. The colored contours indicate thresholds in $\Delta\mathrm{AIC}$ relative to strong and very strong evidence for the inclusion of an additional line, corresponding to $\Delta\mathrm{AIC} = 10$ (purple) and $\Delta\mathrm{AIC} = 20$ (orange), respectively. These conservative $\Delta\mathrm{AIC}$ thresholds are adopted to robustly highlight only the most significant structures; the same regions are also recovered when using likelihood-based $\Delta\chi^2$ ($\Delta W$) contours commonly adopted in previous analyses.}
    \label{fig:fluxscans}
\end{figure}

\begin{figure}
    \centering
    \includegraphics[width=\linewidth]{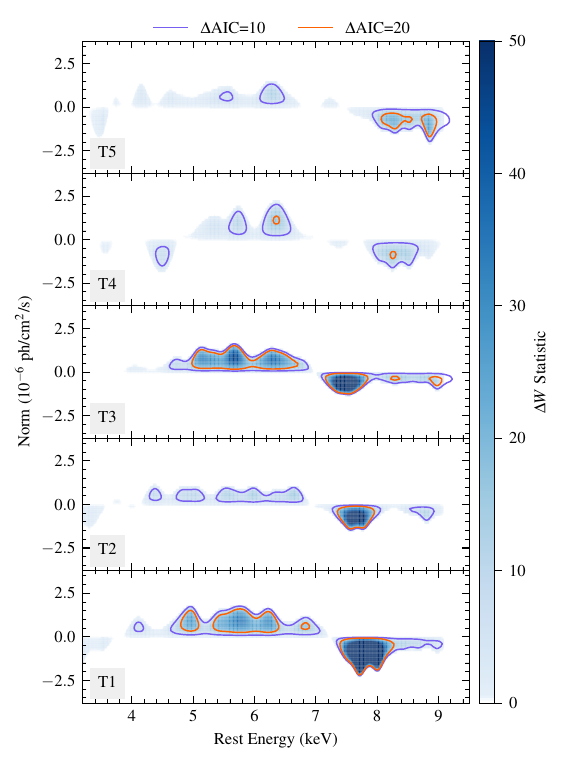}
    \caption{Same as Fig. \ref{fig:fluxscans} but for the time-resolved spectra.}
    \label{fig:timescans}
\end{figure}

\section{Photo-ionization Model}
\label{sec:xstar}
We used the \texttt{XSTARv2.58} photo-ionization code \citep{kallmann_xstar_1999} to generate a grid of photo-ionization absorption spectra, aiming to model the ultra-fast outflow (UFO) in IRAS 13224-3809. The code requires an ionizing spectral energy distribution (SED) and parameters describing the physical properties of the absorbing gas.
For our SED, we adopted a simple power law with a photon index of $\Gamma = 2$ to remain generic, since we cannot favor any of the models A, B, or C based on statistical grounds. To test the impact of a steeper SED, as expected in high accretion regimes, we also generated a grid with $\Gamma = 2.5$. We found no significant differences in the resulting absorption features, except for a systematic shift towards higher ionization parameter values ($\xi$), as anticipated. This shift arises because a steeper power law requires a higher overall luminosity to maintain the same ionizing flux in the hard X-ray band, thereby increasing the ionization parameter. The ionic abundances of Fe XXIV, Fe XXV, and Fe XXVI, those most relevant to the observed absorption, are similarly affected, with the steeper SED causing only a systematic shift in $\xi$ that in our case is of the order of an increase of $\log \xi = 1.1$ (see \citet{luminari_wine_2024} for a deeper discussion about the effect of the SED on X-ray winds that confirms our conclusion for the high ionization potential features in UFOs).
In our model, the column density $N_H$ ranges logarithmically from $10^{22}$ to $10^{25}$ cm$^{-2}$, distributed over 10 steps. $\log\xi$ was varied between 3.0 and 6.0 with steps of 0.2 dex, and the turbulent ranges from 1000 to 30000 km/s with 8 steps. The default solar abundances from \texttt{XSTAR} were applied for the elements.
Around $\log\xi \approx 3.5$, which corresponds to the average value measured for the UFO, the iron ions are distributed with approximately 40\% as Fe XXV and 40\% as Fe XXVI, while the rest are fully ionized. Below this threshold opacity is dominated by Fe XXV while, above it, it is dominated by Fe XXVI. At $\log\xi \sim 4.6$, the gas is almost completely ionized.

All the reported values of column densities in this work are corrected for special relativistic effects using the prescription described in \citet{luminari_wine_2024}.

\section{Additional Materials}
\label{sec:extras}

\begin{figure}
    \centering
    \includegraphics{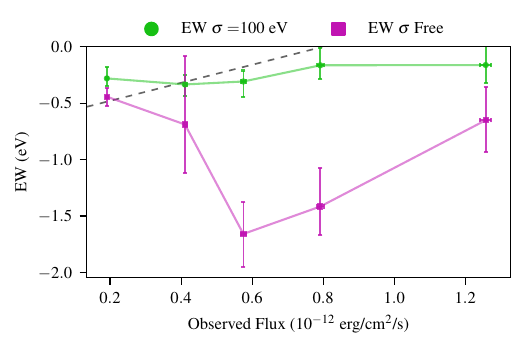}
    \caption{The equivalent width of the UFO as a function of the 3-10 keV observed flux for the time-resolved spectra (T1-T5) of IRAS 13224-3809. Here we are directly visualizing the lack of correlation between the two quantities, in contrast to the findings of \citet{parker_response_2017, pinto_ultrafast_2018}. The dashed line represents the best-fit anti-correlation reported in \citet{parker_response_2017} for comparison.}
    \label{fig:EW-flux}
\end{figure}

\begin{figure}[h!]
    \centering
    \includegraphics{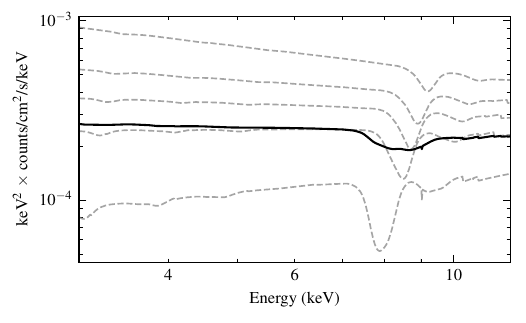}
    \caption{Simulated average spectrum obtained by stacking synthetic spectra generated for each flux-resolved interval, incorporating observed trends in continuum slope, outflow velocity and the tentative ones in ionization and column density. The resulting stacked spectrum (black) is compared to absorbed power laws for each simulated flux interval (dashed gray). The absence of significant emission features in the stacked spectrum indicates that the emission observed in the actual average spectrum is unlikely to be an artifact of stacking variable flat spectra.}
    \label{fig:stacked_simulation}
\end{figure}

\begin{table*}[ht]
\centering
\tiny
\textbf{Phenomenological Model -- Flux-resolved}\\
\vspace{0.5em}
\begin{tabular}{l C{2.5cm} C{2.5cm} C{2.5cm} C{2.5cm} C{2.5cm}}
\toprule
\textbf{Parameter} & \textbf{F1} & \textbf{F2} & \textbf{F3} & \textbf{F4} & \textbf{F5} \\
\midrule
\multicolumn{6}{c}{\textbf{Continuum}} \\
$\Gamma$ & $1.330^{+0.077}_{-0.090}$ & $1.766^{+0.057}_{-0.058}$ & $2.091^{+0.065}_{-0.070}$ & $2.124^{+0.058}_{-0.061}$ & $2.391^{+0.052}_{-0.055}$ \\
$\log_{10} L$ (erg/s) & $42.139^{+0.015}_{-0.011}$ & $42.4642^{+0.0075}_{-0.0074}$ & $42.6067^{+0.0087}_{-0.0081}$ & $42.7367^{+0.0076}_{-0.0072}$ & $42.9214^{+0.0064}_{-0.0061}$ \\
\midrule
\multicolumn{6}{c}{\textbf{Gaussian Absorption 1}} \\
$E$ (keV) & $8.549^{+0.044}_{-0.060}$ & $8.548^{+0.083}_{-0.067}$ & $9.29^{+0.15}_{-0.13}$ & $9.43^{+0.16}_{-0.14}$ & $>9.82$ \\
$\sigma$ (keV) & $0.258^{+0.176}_{-0.095}$ & $0.421^{+0.088}_{-0.078}$ & $1.14^{+0.22}_{-0.17}$ & $1.07^{+0.20}_{-0.16}$ & $0.79^{+0.69}_{-0.38}$ \\
EW (keV) & $-0.443^{+0.079}_{-0.083}$ & $-0.689^{+0.604}_{-0.433}$ & $-1.659^{+0.282}_{-0.289}$ & $-1.416^{+0.345}_{-0.253}$ & $-0.651^{+0.297}_{-0.284}$ \\
\midrule
\multicolumn{6}{c}{\textbf{Gaussian Absorption 2}} \\
$E$ (keV) & $10.04^{+0.38}_{-0.28}$ & $10.02^{+0.38}_{-0.24}$ & - & - & - \\
$\sigma$ (keV) & $1.53^{+0.22}_{-0.52}$ & $1.02^{+0.35}_{-0.32}$ & - & - & - \\
EW (keV) & $-2.606^{+1.004}_{-0.745}$ & $-1.602^{+1.268}_{-1.836}$ & - & - & - \\
\midrule
W-Stat/d.o.f. & $107.913/96=1.12$ & $64.0184/93=0.69$ & $116.929/96=1.22$ & $125.218/98=1.28$ & $105.61/98=1.08$ \\
\bottomrule
\end{tabular}
\vspace{1em}
\caption{Best-fit parameters for the purely phenomenological model (absorbed power law plus broad Gaussian absorption lines) applied to the five flux-resolved spectra (F1–F5). For each fit, the W-Stat/d.o.f. and the reduced statistic are shown. Missing parameters are indicated by $-$.}
\label{tab:flux_phenom_comparison}
\end{table*}

\begin{table*}[ht]
\centering
\tiny
\textbf{Model A -- Flux-resolved}\\
\vspace{0.5em}
\begin{tabular}{l C{2.5cm} C{2.5cm} C{2.5cm} C{2.5cm} C{2.5cm}}
\toprule
\textbf{Parameter} & \textbf{F1} & \textbf{F2} & \textbf{F3} & \textbf{F4} & \textbf{F5} \\
\midrule
\multicolumn{6}{c}{\textbf{Continuum}} \\
$\Gamma$ & $1.54^{+0.13}_{-0.11}$ & $2.09^{+0.28}_{-0.18}$ & $2.218^{+0.050}_{-0.050}$ & $2.221^{+0.048}_{-0.048}$ & $2.437^{+0.122}_{-0.064}$ \\
$\log_{10} L$ (erg/s) & $42.399^{+0.026}_{-0.020}$ & $42.772^{+0.063}_{-0.040}$ & $42.8444^{+0.0083}_{-0.0084}$ & $42.9724^{+0.0076}_{-0.0082}$ & $43.161^{+0.030}_{-0.017}$ \\
\midrule
\multicolumn{6}{c}{\textbf{Ionized Absorber 1}} \\
$N_{\rm H}$ ($10^{23}$ cm$^{-2}$) & $13.3^{+47.8}_{-6.6}$ & $15.4^{+26.3}_{-4.3}$ & $>160$ & $>151$ & $>10.4$ \\
$\log\xi$ & $3.40^{+0.46}_{-0.17}$ & $3.45^{+0.30}_{-0.11}$ & $4.094^{+0.095}_{-0.553}$ & $4.13^{+0.14}_{-0.61}$ & $3.47^{+0.59}_{-0.33}$ \\
$v_{\rm turb}$ (km/s) & $17900^{+4600}_{-4300}$ & $15900^{+2600}_{-2600}$ & $>28000$ & $>27000$ & $>25300$ \\
$v_{\rm out}$ ($c$) & $0.272^{+0.011}_{-0.030}$ & $0.2714^{+0.0073}_{-0.0126}$ & $0.314^{+0.011}_{-0.011}$ & $0.333^{+0.014}_{-0.015}$ & $0.380^{+0.023}_{-0.036}$ \\
$\Delta W / \nu$ & $188.56/4$ & $164.38/4$ & $98.11/4$ & $70.33/4$ & $17.49/4$ \\
$P_0$ & $6.2\times10^{-40}$ & $1.1\times10^{-34}$ & $2.7\times10^{-20}$ & $2.9\times10^{-14}$ & $8.7\times10^{-3}$ \\
\midrule
\multicolumn{6}{c}{\textbf{Ionized Absorber 2}} \\
$N_{\rm H}$ ($10^{23}$ cm$^{-2}$) & $>229$ & $13.4^{+21.7}_{-4.7}$ & - & - & - \\
$\log\xi$ & $4.176^{+0.084}_{-0.496}$ & $3.67^{+0.33}_{-0.17}$ & - & - & - \\
$v_{\rm turb}$ (km/s) & $>27000$ & $>26000$ & - & - & - \\
$v_{\rm out}$ ($c$) & $0.408^{+0.044}_{-0.028}$ & $0.423^{+0.017}_{-0.016}$ & - & - & - \\
$\Delta W / \nu$ & $28.64/4$ & $23.88/4$ & - & - & - \\
$P_0$ & $3.3\times10^{-5}$ & $3.6\times10^{-4}$ & - & - & - \\
\midrule
W-Stat/d.o.f. & $111.395/94=1.19$ & $58.9714/91=0.65$ & $116.398/95=1.23$ & $128.398/97=1.32$ & $104.418/97=1.08$ \\
\bottomrule
\end{tabular}
\vspace{1em}
\caption{Best-fit parameters for Model A applied to the five flux-resolved spectra (F1–F5). For each fit, the W-Stat/d.o.f. and the reduced statistic are shown. Spectra that do not require the second absorber have missing parameters indicated by $-$. The $\Delta W/\nu$ row reports the W-statistic improvement and the number of additional free parameters when adding each absorber component; $P_0 = \exp(-\Delta\mathrm{AIC}/2)$ is the AIC evidence ratio (lower values indicate stronger support for the component).}
\label{tab:flux_modelA_comparison}
\end{table*}

\begin{table*}[ht]
\centering
\tiny

\textbf{Model B -- Flux-resolved}\\
\vspace{0.5em}
\begin{tabular}{l C{2.5cm} C{2.5cm} C{2.5cm} C{2.5cm} C{2.5cm}}
\toprule
\textbf{Parameter} & \textbf{F1} & \textbf{F2} & \textbf{F3} & \textbf{F4} & \textbf{F5} \\
\midrule
\multicolumn{6}{c}{\textbf{Continuum}} \\
$\Gamma$ & $2.02^{+0.10}_{-0.10}$ & $2.274^{+0.139}_{-0.092}$ & $2.279^{+0.103}_{-0.064}$ & $2.779^{+0.070}_{-0.068}$ & $2.684^{+0.048}_{-0.070}$ \\
$\log_{10} L_{\rm{cont}}$ (erg/s) & $42.258^{+0.020}_{-0.020}$ & $42.666^{+0.015}_{-0.033}$ & $42.831^{+0.011}_{-0.031}$ & $42.8670^{+0.0095}_{-0.0096}$ & $43.1068^{+0.0098}_{-0.0072}$ \\
\midrule
\multicolumn{6}{c}{\textbf{Ionized Absorber}} \\
$N_{\rm H}$ ($10^{23}$ cm$^{-2}$) & $>144$ & $>21.1$ & $>164$ & $>205$ & $>26.9$ \\
$\log\xi$ & $4.35^{+0.15}_{-0.46}$ & $3.50^{+0.69}_{-0.17}$ & $4.13^{+0.11}_{-0.57}$ & $4.40^{+0.12}_{-0.84}$ & $3.93^{+0.68}_{-0.39}$ \\
$v_{\rm turb}$ (km/s) & $11200^{+3100}_{-4200}$ & $18600^{+3100}_{-5900}$ & $>28000$ & $>30000$ & $<2000$ \\
$v_{\rm out}$ ($c$) & $0.2502^{+0.0058}_{-0.0048}$ & $0.2734^{+0.0079}_{-0.0092}$ & $0.318^{+0.012}_{-0.012}$ & $0.345^{+0.021}_{-0.030}$ & $0.3436^{+0.0065}_{-0.0053}$ \\
$\Delta W / \nu$ & $38.42/4$ & $48.75/4$ & $22.20/4$ & $6.45/4$ & $7.33/4$ \\
$P_0$ & $2.5\times10^{-7}$ & $1.4\times10^{-9}$ & $8.2\times10^{-4}$ & $>0.05$ & $>0.05$ \\
\midrule
\multicolumn{6}{c}{\textbf{Broad Gaussian Emission}} \\
$E$ (keV) & $6.44^{+0.16}_{-0.14}$ & $6.88^{+0.11}_{-0.23}$ & $6.03^{+0.44}_{-0.28}$ & $6.35^{+0.11}_{-0.11}$ & $6.74^{+0.21}_{-0.21}$ \\
$\sigma$ (keV) & $>1.0$ & $>0.83$ & $0.50^{+0.97}_{-0.50}$ & $>0.87$ & $>0.64$ \\
EW (keV) & $1.071^{+0.349}_{-0.286}$ & $0.677^{+0.467}_{-0.293}$ & $0.102^{+0.217}_{-0.085}$ & $1.244^{+0.258}_{-0.222}$ & $0.601^{+0.331}_{-0.237}$ \\
$\log_{10} L_{\rm{gauss}}$ (erg/s) & $41.437^{+0.066}_{-0.082}$ & $41.624^{+0.088}_{-0.122}$ & $41.00^{+0.57}_{-0.42}$ & $42.041^{+0.047}_{-0.054}$ & $41.932^{+0.087}_{-0.157}$ \\
$\Delta W / \nu$ & $160.38/3$ & $129.52/3$ & $78.38/3$ & $76.14/3$ & $18.45/3$ \\
$P_0$ & $3.0\times10^{-34}$ & $1.5\times10^{-27}$ & $1.9\times10^{-16}$ & $5.9\times10^{-16}$ & $2.0\times10^{-3}$ \\
\midrule
W-Stat/d.o.f. & $129.79/95=1.37$ & $68.9687/92=0.75$ & $113.927/92=1.24$ & $122.589/98=1.25$ & $103.45/98=1.06$ \\
\bottomrule
\end{tabular}
\vspace{1em}
\caption{Best-fit parameters for Model B applied to the five flux-resolved spectra (F1–F5). The Ionized Absorber section is shown where required by the fit. The Broad Gaussian Emission section includes centroid, width, equivalent width, and luminosity. The continuum section reports the photon index and continuum luminosity. For each fit, the W-Stat/d.o.f. and the reduced statistic are shown. Missing parameters are indicated by $-$. The $\Delta W/\nu$ and $P_0$ rows report the fit improvement and AIC evidence ratio for the Gaussian emission (vs.~bare power law) and for the absorber (vs.~power law + Gaussian); see Table~\ref{tab:flux_modelA_comparison} caption for details.}
\label{tab:flux_modelB_comparison}
\end{table*}

\begin{table*}[ht]
\centering
\tiny
\textbf{Model C -- Flux-resolved}\\
\vspace{0.5em}
\begin{tabular}{l C{2.5cm} C{2.5cm} C{2.5cm} C{2.5cm} C{2.5cm}}
\toprule
\textbf{Parameter} & \textbf{F1} & \textbf{F2} & \textbf{F3} & \textbf{F4} & \textbf{F5} \\
\midrule
\multicolumn{6}{c}{\textbf{Relativistic Reflection}} \\
$\log\xi$ & $3.657^{+0.085}_{-0.126}$ & $3.716^{+0.069}_{-0.105}$ & $4.19^{+0.12}_{-0.12}$ & $4.084^{+0.076}_{-0.072}$ & $4.09^{+0.34}_{-0.40}$ \\
$\Gamma$ & $2.132^{+0.077}_{-0.104}$ & $2.153^{+0.128}_{-0.074}$ & $2.220^{+0.088}_{-0.049}$ & $2.399^{+0.087}_{-0.042}$ & $2.495^{+0.077}_{-0.116}$ \\
$\log_{10} L_{\rm{cont}}$ (erg/s) & $42.391^{+0.020}_{-0.010}$ & $42.6748^{+0.0149}_{-0.0086}$ & $42.8127^{+0.0082}_{-0.0072}$ & $42.9277^{+0.0053}_{-0.0055}$ & $43.1330^{+0.0055}_{-0.0055}$ \\
$R_{\rm refl}$ & $>90$ & $>85$ & $>44$ & $>18$ & $1.33^{+5.67}_{-0.61}$ \\
$\Delta W / \nu$ & $164.37/2$ & $143.63/2$ & $80.54/2$ & $72.68/2$ & $17.23/2$ \\
$P_0$ & $1.5\times10^{-35}$ & $4.8\times10^{-31}$ & $2.4\times10^{-17}$ & $1.2\times10^{-15}$ & $1.3\times10^{-3}$ \\
\midrule
\multicolumn{6}{c}{\textbf{Ionized Absorber}} \\
$N_{\rm H}$ ($10^{23}$ cm$^{-2}$) & $4.80^{+1.23}_{-0.80}$ & $12.0^{+5.3}_{-2.0}$ & $>142$ & $4.01^{+3.4}_{-1.5}$ & $24^{+26}_{-20}$ \\
$\log\xi$ & $<3.01$ & $3.470^{+0.045}_{-0.187}$ & $4.33^{+0.18}_{-0.13}$ & $3.60^{+1.43}_{-0.11}$ & $4.11^{+1.03}_{-0.14}$ \\
$v_{\rm turb}$ (km/s) & $21100^{+4000}_{-3500}$ & $14000^{+4500}_{-3200}$ & $22200^{+6300}_{-5000}$ & $>8000$ & $<4000$ \\
$v_{\rm out}$ ($c$) & $0.2923^{+0.0077}_{-0.0075}$ & $0.2685^{+0.0087}_{-0.0083}$ & $0.289^{+0.015}_{-0.013}$ & $0.293^{+0.032}_{-0.048}$ & $0.3435^{+0.034}_{-0.023}$ \\
$\Delta W / \nu$ & $52.26/4$ & $45.25/4$ & $23.16/4$ & $13.60/4$ & $6.80/4$ \\
$P_0$ & $2.5\times10^{-10}$ & $8.2\times10^{-9}$ & $5.1\times10^{-4}$ & $>0.05$ & $>0.05$ \\
\midrule
W-Stat/d.o.f. & $111.966/96=1.17$ & $58.3485/93=0.63$ & $110.809/93=1.19$ & $126.047/99=1.27$ & $104.672/99=1.06$ \\
\bottomrule
\end{tabular}
\vspace{1em}
\caption{Best-fit parameters for Model C (relxill) applied to the five flux-resolved spectra (F1–F5). Ionized Absorber parameters are shown where required by the fit. The relativistic reflection section includes spin ($a$), inclination ($i$), ionization ($\log\xi$), iron abundance ($A_{\rm Fe}$), photon index ($\Gamma$), continuum luminosity, and reflection fraction. For each fit, the W-Stat/d.o.f. and the reduced statistic are shown. Missing parameters are indicated by $-$. The $\Delta W/\nu$ and $P_0$ rows report the fit improvement and AIC evidence ratio for the \texttt{relxill} component (vs.~bare power law) and for the absorber (vs.~\texttt{relxill} alone); see Table~\ref{tab:flux_modelA_comparison} caption for details.}
\label{tab:flux_modelC_comparison}
\end{table*}

\begin{table*}[ht]
\centering
\tiny
\textbf{Phenomenological Model -- Time-resolved}\\
\vspace{0.5em}
\begin{tabular}{l C{2.5cm} C{2.5cm} C{2.5cm} C{2.5cm} C{2.5cm}}
\toprule
\textbf{Parameter} & \textbf{T1} & \textbf{T2} & \textbf{T3} & \textbf{T4} & \textbf{T5} \\
\midrule
\multicolumn{6}{c}{\textbf{Continuum}} \\
$\Gamma$ & $1.735^{+0.066}_{-0.074}$ & $1.856^{+0.061}_{-0.062}$ & $1.514^{+0.060}_{-0.065}$ & $2.220^{+0.064}_{-0.165}$ & $2.287^{+0.056}_{-0.060}$ \\
$\log_{10} L$ (erg/s) & $42.6500^{+0.0104}_{-0.0088}$ & $42.5861 \pm 0.0078$ & $42.5360^{+0.0093}_{-0.0081}$ & $42.8816^{+0.0458}_{-0.0087}$ & $42.8374^{+0.0072}_{-0.0068}$ \\
\midrule
\multicolumn{6}{c}{\textbf{Gaussian Absorption 1}} \\
$E$ (keV) & $8.938^{+0.057}_{-0.055}$ & $8.662^{+0.063}_{-0.068}$ & $8.412 \pm 0.045$ & $9.41^{+0.18}_{-0.17}$ & $9.78^{+0.37}_{-0.20}$ \\
$\sigma$ (keV) & $0.321^{+0.097}_{-0.079}$ & $0.442^{+0.091}_{-0.082}$ & $0.325^{+0.074}_{-0.061}$ & $0.66^{+0.25}_{-0.20}$ & $0.87^{+0.43}_{-0.25}$ \\
EW (keV) & $-0.58^{+0.14}_{-0.11}$ & $-0.73^{+0.10}_{-0.11}$ & $-0.54^{+0.16}_{-0.13}$ & $-0.95^{+0.20}_{-0.25}$ & $-1.37^{+0.42}_{-0.42}$ \\
\midrule
\multicolumn{6}{c}{\textbf{Gaussian Absorption 2}} \\
$E$ (keV) & $>10$ & $10.028^{+0.099}_{-0.077}$ & $10.05^{+0.33}_{-0.22}$ & $>10$ & - \\
$\sigma$ (keV) & $1.31^{+0.41}_{-0.45}$ & $0.32^{+0.25}_{-0.17}$ & $1.17^{+0.35}_{-0.32}$ & $0.174^{+0.557}_{-0.064}$ & - \\
EW (keV) & $-1.95^{+0.67}_{-0.69}$ & $-0.56^{+0.16}_{-0.10}$ & $-1.95^{+0.97}_{-0.92}$ & $-0.100^{+0.030}_{-0.025}$ & - \\
\midrule
W-Stat/d.o.f. & $77.2/69=1.12$ & $65.7/69=0.95$ & $73.4/73=1.01$ & $66.2/73=0.91$ & $71.6/78=0.92$ \\
\bottomrule
\end{tabular}
\vspace{1em}
\caption{Same as Table~\ref{tab:flux_phenom_comparison}, but for the five time-resolved spectra (T1--T5).}
\label{tab:time_phenom_comparison}
\end{table*}

\begin{table*}[ht]
\centering
\tiny
\textbf{Model A -- Time-resolved}\\
\vspace{0.5em}
\begin{tabular}{l C{2.5cm} C{2.5cm} C{2.5cm} C{2.5cm} C{2.5cm}}
\toprule
\textbf{Parameter} & \textbf{T1} & \textbf{T2} & \textbf{T3} & \textbf{T4} & \textbf{T5} \\
\midrule
\multicolumn{6}{c}{\textbf{Continuum}} \\
$\Gamma$ & $2.47^{+0.30}_{-0.37}$ & $1.942^{+0.058}_{-0.059}$ & $1.75^{+0.11}_{-0.12}$ & $2.290^{+0.053}_{-0.052}$ & $2.361^{+0.046}_{-0.046}$ \\
$\log_{10} L$ (erg/s) & $42.824 \pm 0.059$ & $42.6180^{+0.0094}_{-0.0096}$ & $42.602^{+0.019}_{-0.017}$ & $42.8964 \pm 0.0084$ & $42.8714^{+0.0267}_{-0.0081}$ \\
\midrule
\multicolumn{6}{c}{\textbf{Ionized Absorber 1}} \\
$N_{\rm H}$ ($10^{23}$ cm$^{-2}$) & $11.4^{+14.1}_{-4.2}$ & $>137$ & $10.4^{+11.5}_{-3.7}$ & $>201$ & $>165$ \\
$\log\xi$ & $3.29^{+0.27}_{-0.16}$ & $4.14^{+0.11}_{-0.33}$ & $3.34^{+0.35}_{-0.15}$ & $4.374^{+0.072}_{-1.103}$ & $4.14^{+0.16}_{-0.69}$ \\
$v_{\rm turb}$ (km/s) & $13500^{+3800}_{-3200}$ & $20600^{+4100}_{-2900}$ & $15900^{+3000}_{-2400}$ & $>25700$ & $>29500$ \\
$v_{\rm out}$ ($c$) & $0.311^{+0.010}_{-0.018}$ & $0.264^{+0.009}_{-0.007}$ & $0.248^{+0.009}_{-0.017}$ & $0.349^{+0.015}_{-0.019}$ & $0.368^{+0.016}_{-0.015}$ \\
$\Delta W / \nu$ & $176.51/4$ & $89.12/4$ & $188.25/4$ & $28.36/4$ & $54.19/4$ \\
$P_0$ & $2.6\times10^{-37}$ & $2.4\times10^{-18}$ & $7.2\times10^{-40}$ & $3.8\times10^{-5}$ & $9.3\times10^{-11}$ \\
\midrule
\multicolumn{6}{c}{\textbf{Ionized Absorber 2}} \\
$N_{\rm H}$ ($10^{23}$ cm$^{-2}$) & $7.5^{+8.3}_{-1.8}$ & - & $>220$ & - & - \\
$\log\xi$ & $<3.04$ & - & $4.19^{+0.08}_{-0.65}$ & - & - \\
$v_{\rm turb}$ (km/s) & $>25400$ & - & $>29500$ & - & - \\
$v_{\rm out}$ ($c$) & $0.422^{+0.012}_{-0.017}$ & - & $0.389^{+0.031}_{-0.021}$ & - & - \\
$\Delta W / \nu$ & $18.95/4$ & - & $32.46/4$ & - & - \\
$P_0$ & $4.2\times10^{-3}$ & - & $4.9\times10^{-6}$ & - & - \\
\midrule
W-Stat/d.o.f. & $115.7/91=1.27$ & $101.647/90=1.13$ & $89.2986/90=0.99$ & $98.8593/92=1.07$ & $102.579/94=1.09$ \\
\bottomrule
\end{tabular}
\vspace{1em}
\caption{Same as Table~\ref{tab:flux_modelA_comparison}, but for the five time-resolved spectra (T1--T5).}
\label{tab:time_modelA_comparison}
\end{table*}

\begin{table*}[ht]
\centering
\tiny
\textbf{Model B -- Time-resolved}\\
\vspace{0.5em}
\begin{tabular}{l C{2.5cm} C{2.5cm} C{2.5cm} C{2.5cm} C{2.5cm}}
\toprule
\textbf{Parameter} & \textbf{T1} & \textbf{T2} & \textbf{T3} & \textbf{T4} & \textbf{T5} \\
\midrule
\multicolumn{6}{c}{\textbf{Continuum}} \\
$\Gamma$ & $2.221^{+0.104}_{-0.061}$ & $2.20^{+0.11}_{-0.15}$ & $2.45^{+0.19}_{-0.14}$ & $2.592^{+0.095}_{-0.085}$ & $2.391^{+0.110}_{-0.061}$ \\
$\log_{10} L_{\rm{cont}}$ (erg/s) & $42.615^{+0.019}_{-0.020}$ & $42.566^{+0.032}_{-0.021}$ & $42.581^{+0.067}_{-0.064}$ & $42.828^{+0.012}_{-0.013}$ & $42.866^{+0.012}_{-0.023}$ \\
\midrule
\multicolumn{6}{c}{\textbf{Ionized Absorber}} \\
$N_{\rm H}$ ($10^{23}$ cm$^{-2}$) & $>68$ & $>46$ & $5.8^{+2.1}_{-1.9}$ & $7.3 \pm 0.4$ & $>63$ \\
$\log\xi$ & $4.13^{+0.17}_{-0.57}$ & $3.94^{+0.54}_{-0.35}$ & $3.24^{+0.24}_{-0.15}$ & $3.5^{+0.20}_{-0.40}$ & $4.05^{+0.30}_{-0.70}$ \\
$v_{\rm turb}$ (km/s) & $14000^{+4600}_{-2500}$ & $12600^{+8700}_{-6800}$ & $19500^{+4500}_{-5900}$ & $2450^{-500}_{-2350}$ & $>25000$ \\
$v_{\rm out}$ ($c$) & $0.298^{+0.006}_{-0.006}$ & $0.263^{+0.009}_{-0.008}$ & $0.281^{+0.007}_{-0.008}$ & $0.344^{+0.019}_{-0.015}$ & $0.371^{+0.022}_{-0.016}$ \\
$\Delta W / \nu$ & $52.72/4$ & $28.87/4$ & $47.10/4$ & $14.01/4$ & $20.94/4$ \\
$P_0$ & $2.0\times10^{-10}$ & $2.9\times10^{-5}$ & $3.2\times10^{-9}$ & $0.048$ & $1.6\times10^{-3}$ \\
\midrule
\multicolumn{6}{c}{\textbf{Broad Gaussian Emission}} \\
$E$ (keV) & $6.39^{+0.22}_{-0.21}$ & $6.25^{+0.41}_{-0.36}$ & $6.67^{+0.26}_{-0.10}$ & $6.73^{+0.15}_{-0.16}$ & $6.76 \pm 0.68$ \\
$\sigma$ (keV) & $>1.0$ & $>0.87$ & $>0.92$ & $0.76 \pm 0.22$ & $0.50^{+1.10}_{-0.50}$ \\
EW (keV) & $0.689^{+0.224}_{-0.422}$ & $0.471^{+0.226}_{-0.292}$ & $1.059^{+0.451}_{-0.273}$ & $0.893^{+0.251}_{-0.187}$ & $0.052^{+0.148}_{-0.052}$ \\
$\log_{10} L_{\rm{gauss}}$ (erg/s) & $41.59^{+0.10}_{-0.13}$ & $41.39 \pm 0.14$ & $41.682^{+0.064}_{-0.068}$ & $41.83^{+0.10}_{-0.12}$ & $40.68^{+0.73}_{-40.68}$ \\
$\Delta W / \nu$ & $131.83/3$ & $64.49/3$ & $172.05/3$ & $38.18/3$ & $33.78/3$ \\
$P_0$ & $4.8\times10^{-28}$ & $2.0\times10^{-13}$ & $8.8\times10^{-37}$ & $1.0\times10^{-7}$ & $9.3\times10^{-7}$ \\
\midrule
W-Stat/d.o.f. & $126.612/92=1.38$ & $97.4065/87=1.12$ & $90.8694/91=1.00$ & $89.0397/93=0.96$ & $102.045/91=1.12$ \\
\bottomrule
\end{tabular}
\vspace{1em}
\caption{Same as Table~\ref{tab:flux_modelB_comparison}, but for the five time-resolved spectra (T1--T5).}
\label{tab:time_modelB_comparison}
\end{table*}

\begin{table*}[ht]
\centering
\tiny
\textbf{Model C -- Time-resolved}\\
\vspace{0.5em}
\begin{tabular}{l C{2.5cm} C{2.5cm} C{2.5cm} C{2.5cm} C{2.5cm}}
\toprule
\textbf{Parameter} & \textbf{T1} & \textbf{T2} & \textbf{T3} & \textbf{T4} & \textbf{T5} \\
\midrule
\multicolumn{6}{c}{\textbf{Relativistic Reflection}} \\
$\log\xi$ & $3.725^{+0.079}_{-0.059}$ & $3.86^{+0.12}_{-0.28}$ & $3.804^{+0.100}_{-0.145}$ & $4.01^{+0.18}_{-0.22}$ & $4.315^{+0.088}_{-0.070}$ \\
$\Gamma$ & $2.000^{+0.062}_{-0.041}$ & $1.992^{+0.114}_{-0.066}$ & $2.350^{+0.075}_{-0.138}$ & $2.445^{+0.091}_{-0.078}$ & $2.408^{+0.095}_{-0.044}$ \\
$\log_{10} L_{\rm{cont}}$ (erg/s) & $42.6492^{+0.0084}_{-0.0064}$ & $42.5905^{+0.0083}_{-0.0087}$ & $42.5948^{+0.0193}_{-0.0087}$ & $42.8679^{+0.0059}_{-0.0060}$ & $42.8332^{+0.0052}_{-0.0053}$ \\
$R_{\rm refl}$ & $>55$ & $>73$ & $>25$ & $3.1^{+10.1}_{-1.3}$ & $>12$ \\
$\Delta W / \nu$ & $130.49/2$ & $71.99/2$ & $165.22/2$ & $33.71/2$ & $46.80/2$ \\
$P_0$ & $3.4\times10^{-28}$ & $1.7\times10^{-15}$ & $9.8\times10^{-36}$ & $3.5\times10^{-7}$ & $5.1\times10^{-10}$ \\
\midrule
\multicolumn{6}{c}{\textbf{Ionized Absorber}} \\
$N_{\rm H}$ ($10^{23}$ cm$^{-2}$) & $55^{+105}_{-36}$ & $36^{+34}_{-13}$ & $4.57^{+1.24}_{-0.28}$ & $7.9^{+31.7}_{-4.4}$ & $27^{+15}_{-19}$ \\
$\log\xi$ & $3.90^{+0.59}_{-0.32}$ & $3.90^{+0.72}_{-0.28}$ & $3.22^{+0.21}_{-0.18}$ & $3.58^{+0.43}_{-0.19}$ & $3.64^{+0.07}_{-0.53}$ \\
$v_{\rm turb}$ (km/s) & $12900^{+2800}_{-2200}$ & $13100^{+6600}_{-3600}$ & $20300^{+3600}_{-3300}$ & $<3200$ & $>25000$ \\
$v_{\rm out}$ ($c$) & $0.288^{+0.006}_{-0.005}$ & $0.259^{+0.007}_{-0.009}$ & $0.273^{+0.009}_{-0.010}$ & $0.3429^{+0.016}_{-0.018}$ & $0.388^{+0.032}_{-0.023}$ \\
$\Delta W / \nu$ & $61.32/4$ & $27.32/4$ & $60.17/4$ & $9.80/4$ & $8.55/4$ \\
$P_0$ & $2.6\times10^{-12}$ & $6.4\times10^{-5}$ & $4.7\times10^{-12}$ & $>0.05$ & $>0.05$ \\
\midrule
W-Stat/d.o.f. & $119.355/93=1.28$ & $91.4517/88=1.04$ & $84.6258/92=0.92$ & $93.507/94=0.99$ & $109.97/96=1.15$ \\
\bottomrule
\end{tabular}
\vspace{1em}
\caption{Same as Table~\ref{tab:flux_modelC_comparison}, but for the five time-resolved spectra (T1--T5).}
\label{tab:time_modelC_comparison}
\end{table*}

\begin{figure*}
    \centering
    \includegraphics[width=\linewidth]{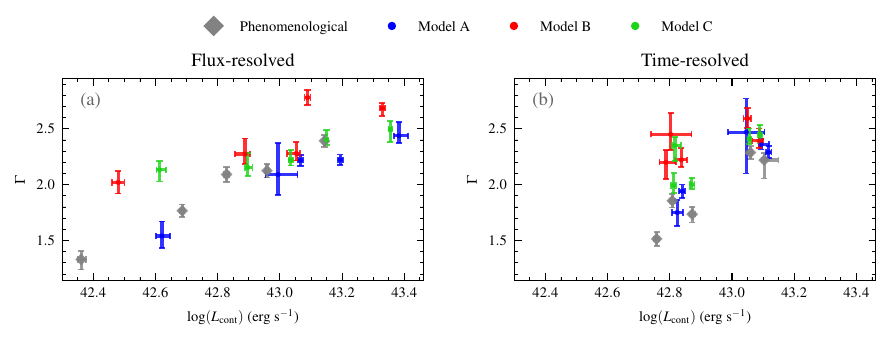}
    \includegraphics[width=\linewidth]{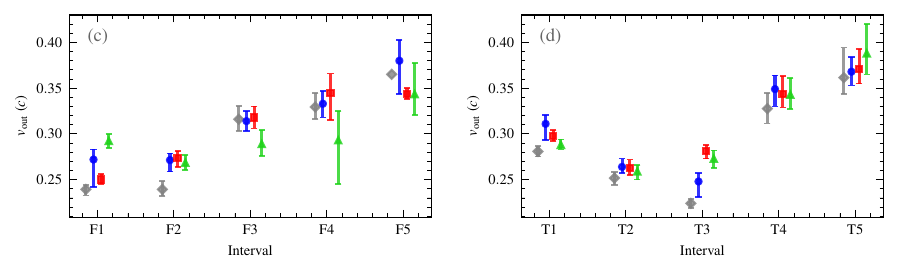}
    \includegraphics[width=\linewidth]{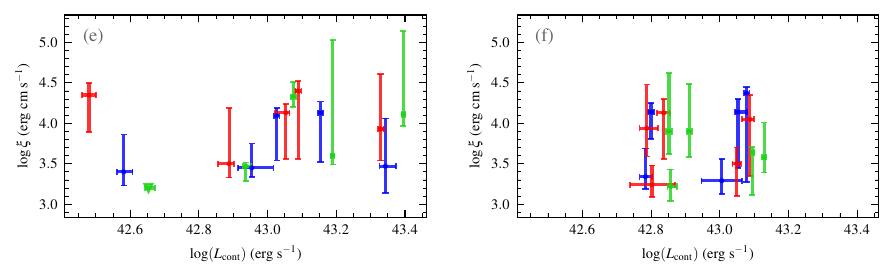}
    \includegraphics[width=\linewidth]{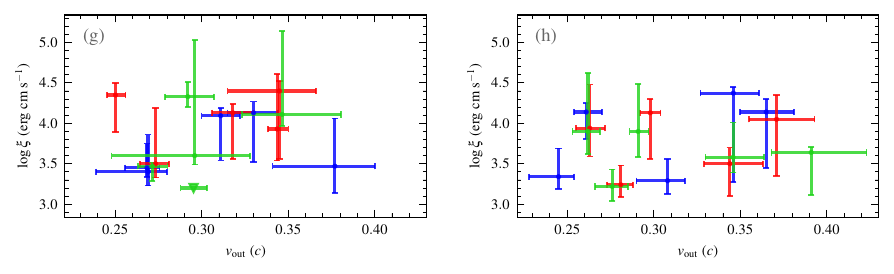}
    \caption{Visualize the key findings in Flux- and Time-Resolved Analysis. Missing values are reported as $\times$ and upper--lower limits as triangles. Panles c) and d) show the outflow velocities per each flux- and time-resolved spectrum, respectively showing clearly the consistency of $v_{\rm out}$ across the models. For better visibility, values from each model are offset by a small amount in the x-axis.}
    \label{fig:FLUX_TIME_recap}
\end{figure*}

\end{document}